\documentclass{aa}
\usepackage{txfonts}
\usepackage{graphicx}
\usepackage{natbib}
\usepackage{longtable,lscape}
\bibpunct{(}{)}{;}{a}{}{,}

\begin{document}
   \title{An abundance study of red-giant-branch stars in the Hercules dwarf spheroidal galaxy}


\author{ D.\,Ad\'{e}n\inst{1} \and K.\,Eriksson\inst{2} \and S.\,Feltzing\inst{1} \and E. K. Grebel\inst{3} \and A. Koch\inst{4} \and M. I. Wilkinson\inst{4}}

\authorrunning{Ad\'en et al.}

\offprints{D. Ad\'{e}n  daniel.aden@astro.lu.se }
 
\institute{Lund Observatory, Box 43, SE-22100 Lund, Sweden \and
 Department of Physics and Astronomy, Uppsala University, Box 515, SE-751 20 Uppsala, Sweden \and
Astronomisches Rechen-Institut, Zentrum f\"ur Astronomie der 
Universit\"at Heidelberg, M\"onchhofstr. 12-14, 69120 Heidelberg, Germany \and  Department of Physics and Astronomy, University of Leicester, University Road, Leicester LE1 7RH, UK  }

   \date{Received 10 May 2010 /
   Accepted 24 October 2010}

 
  \abstract
   {Dwarf spheroidal galaxies are some of the most metal-poor, and least luminous objects known. Detailed elemental
   abundance analysis of stars in these faint objects is key to our understanding of star formation and chemical
   enrichment in the early universe, and may provide useful information on how larger galaxies form.
   }
   {Our aim is to provide a determination of [Fe/H] and [Ca/H] for confirmed red-giant branch member stars of the Hercules
   dwarf spheroidal galaxy. Based on this we explore the ages of the prevailing stellar populations in
   Hercules, and the enrichment history from supernovae. Additionally, we aim to provide a
   new simple metallicity calibration for Str\"omgren photometry for metal-poor, red giant branch stars.
   }
   {High-resolution, multi-fibre spectroscopy and Str\"omgren photometry are combined to provide
   as much information on the stars as possible. From this we derive abundances by solving the radiative transfer
   equations through {\sc marcs} {\rm model atmospheres.}
   }
   {We find that the red-giant branch stars of the Hercules dSph galaxy are more
   metal-poor than estimated in our previous study that was based on photometry alone.
   From this, we derive a new metallicity calibration for the Str\"omgren photometry.
   Additionally, we find an abundance trend such that [Ca/Fe] is higher for more metal-poor stars,
   and lower for more metal-rich stars, with a spread of about 0.8 dex. 
   The [Ca/Fe] trend suggests an early rapid chemical enrichment through
   supernovae of type II, followed by a phase of slow star formation 
   dominated by enrichment through supernovae of type Ia.  A comparison
   with isochrones indicates that the red giants in Hercules are older than 10 Gyr.   
   }
   {}

\keywords{ Galaxies:dwarf -- 
		Galaxies: evolution --
		Galaxies: individual: Hercules --
		Stars: abundances
		}
		
   \maketitle
%

\section{Introduction}

Over the past few years, the number of known dwarf spheroidal (dSph) galaxies orbiting
the Milky Way has more than doubled through systematic searches in large photometric surveys such
as the Sloan Digital Sky Survey
\citep[e.g.,][]{2006ApJ...650L..41Z,2007ApJ...654..897B}.  These
recently discovered dSph galaxies have much lower surface brightness than the
previously known dSph galaxies. Typically, these systems have total luminosities
$M_V > -6$ \citep[e.g.,][]{2008ApJ...684.1075M}, so
they are named ultra-faint \citep[e.g.,][]{2009AN....330..675K}.
Additionally, they are so faint that thus far they have only been detected around
the Milky Way, but more sensitive surveys in the future
may yield additional detections also at larger distances \citep{2008ApJ...688..277T}.
These galaxies are also more metal-poor than the classical, more luminous dSph galaxies \citep[e.g.,][]{2007ApJ...670..313S,2008ApJ...685L..43K}

Very metal-poor stars have been found in the Galactic halo
\citep[e.g., ][]{2009A&A...507..817S}.
However, studies of the more luminous dSph galaxies
\citep[e.g.,][]{2006AJ....131..895K,2006ApJ...651L.121H} have found a significant lack of stars with
${\rm [Fe/H]} \leq -3$ when compared to the stars of the Milky Way
halo. Since dSph galaxies may be the building blocks
of parts of the Galactic halo \citep[see][for a discussion of the ages of stellar populations in the Galactic dSphs and halo]{2004ApJ...610L..89G}, this result was considered a problem for 
our current understanding of the formation of large galaxies \citep{2009A&A...507..817S}.
However, recent studies of the ultra-faint \citep[e.g.,][]{2008ApJ...685L..43K,2010ApJ...708..560F,2010ApJ...711..350N,2010arXiv1001.3137S}
and the classical \citep[e.g.,][]{2010A&A...513A..34S,2010Natur.464...72F}
dSph galaxies have discovered stars with ${\rm [Fe/H]} \leq -3$, thus
reigniting the discussion of the origin of the Galactic halo. Whether the metallicity distribution
functions of the halo and of the dSph galaxies agree still remain to be determined. Additional abundance
studies are needed for both halo and dSph stars, in order to rule out selection biases due to low number statistics.

Studies of ${\rm [\alpha/Fe]}$ in the stars in the more luminous dSph galaxies
suggest that stars more metal-rich than ${\rm [Fe/H] > -2 }$ have lower ${\rm [\alpha/Fe]}$ ratios,
whilst more metal-poor stars (from now taken to be stars with ${\rm [Fe/H] < -2 }$) have about the same enhancement in the $\alpha$-elements relative to iron
as the stars in the halo and
disk of the Milky Way \citep[see, e.g.,][]{2001ApJ...548..592S,2007PASP..119..939G,2009ARA&A..47..371T}.
This could be interpreted as support for an early accretion of dSphs.
The discrepancy is poorly constrained for
the recently discovered ultra-faint dSph galaxies,
with notable exceptions for a few stars in these ultra-faint objects \citep[][]{2009A&A...508L...1F,2010ApJ...708..560F}.

The Hercules dSph galaxy lies at a distance of $\sim 150$ kpc from us
\citep{2009A&A...506.1147A} and it has a $V$-band surface
brightness of only ${\rm 27.2 \pm 0.6 \, mag\, arcsec^{-2}}$
\citep{2008ApJ...684.1075M}. Previous studies, based on photometry
and the measurements of the near-infrared Ca\,{\sc ii} triplet lines in red giant
branch (RGB) stars, have found a mean metallicity of ${\rm [Fe/H] \sim
-2.3}$ \citep{2007ApJ...670..313S,2009A&A...506.1147A}. 
In Table \ref{table:prop} we provide a list with additional properties of the Hercules dSph galaxy.

A study
using spectrum synthesis of Fe\,{\sc i} lines \citep{2008ApJ...685L..43K}
found a lower mean metallicity of $-2.58 \pm 0.51$ dex.
\citet{2008ApJ...688L..13K} found, using high-resolution spectroscopy
of two Hercules RGB stars, that the hydrostatic burning $\alpha$-elements (e.g., Mg, O)
are strongly enhanced, while the heavy (mainly) s-process elements (e.g., Y,
Sr, Ba, La) are largely depleted. 
The low [Fe/H] observed for the Hercules dSph galaxy suggests that
star formation ceased relatively early after the formation of this galaxy.
Thus, detailed elemental abundances for stars in the ultra-faint dSph galaxies
are key to our understanding of star formation and chemical enrichment in the early universe.

In this study we will determine some of the elemental abundance trends
in the ultra-faint Hercules dSph galaxy.

\begin{table}
\caption{Properties of the Hercules dSph galaxy}
\label{table:prop}
\centering
\begin{tabular}{c c c c}
\hline\hline
 Parameter & & & Footnote \\
\hline
$\alpha_0$ & J2000 & 16 31 05.2 $\pm$ 2.5 & a \\  
$\delta_0$  & J2000 & + 12 47 18 $\pm$ 17 & a \\
$r_h$ & arcmin & $8.6^{+1.8}_{-1.1}$ & a \\
$M_V$ & & $-6.6\pm 0.3$ & a \\
D & kpc & $147^{+8}_{-7}$ & b \\
$E(B-V)$ &  & 0.062 & b \\
$M_{300}$ & $M_{\odot}$ & $1.9^{+1.1}_{-0.8} \cdot 10^6$ & c \\
\hline
\end{tabular}
\begin{list}{}{}
\item[a] The centroid, $\alpha_0$ and $\delta_0$, half-light radius, $r_h$, and absolute magnitude
are taken from \citet{2008ApJ...684.1075M}.
\item[b] The distance, $D$, and reddening, $E(B-V)$, are taken from  \citet{2009A&A...506.1147A} 
\item[c] The mass within the central 300 pc is taken from \citet{2009ApJ...706L.150A}.
\end{list}
\end{table}

This paper is organised as follows: in Sect. \ref{obs} we describe the
observations and the reduction of our spectra. In
Sect. \ref{stepar} we describe the determination of the stellar
parameters for each star. Section \ref{abund} deals with the abundance
analysis. In
Sect. \ref{comparison} we provide a comparison with abundances
determined in other studies, in Sect. \ref{discu} we show and discuss our
results and Sect. \ref{conclusion} concludes the article.

\section{Observations, data reduction, and measurement of equivalent widths} \label{obs}

\subsection{Selection of targets}

Some of the new ultra-faint dSph galaxies are seen through a significant portion
of the Milky Way disk. Moreover, sometimes they have systemic velocities
very similar to the bulk motion of the stars in the Milky Way disk. This is the case for
the Hercules dSph galaxy \citep{2009A&A...506.1147A}.
Thus, when studying systems like Hercules it is very important that the stars
are confirmed members of the galaxy, and not foreground contaminating stars that
belong to the Milky Way.  In \citet{2009A&A...506.1147A} we showed
that the mean velocity of the Hercules dSph is very similar to the velocity distribution
 of the foreground dwarf stars, making it difficult
to disentangle the dSph galaxy stars from the foreground dwarf stars
using radial velocity measurements alone.  We used the Str\"omgren
$c_1$ index to identify the RGB stars that belong
to the dSph galaxy and
showed that a proper cleaning of the sample results in a smaller value for the 
velocity dispersion of the system. This has implications for galaxy properties derived from such 
velocity dispersions, e.g., resulting in a lower mass \citep{2009ApJ...706L.150A,2009ApJ...704.1274W}.
In this study, we revisit the previously identified RGB
stars of the Hercules dSph galaxy with high-resolution
spectroscopy.

The RGB stars for this study were taken from the list of Hercules
dSph galaxy members in \citet{2009A&A...506.1147A}.  We selected RGB
stars brighter than $V_0 \sim 20$ (see Fig. \ref{iso}).  Stars fainter
than $V_0 \sim 20$ were not considered since the signal-to-noise ratio, per pixel,
(S/N) would be too low for equivalent width measurements. In total,
20 RGB stars were selected (see Table \ref{table_obs}).

\begin{table}
\caption{Data for the RGB stars in the Hercules dSph galaxy observed with FLAMES.}
\label{table_obs}
\centering
\begin{tabular}{c c c c c c c}
\hline\hline
ID & RA & DEC & $V$ & $(b-y)$ & S/N & Used \\
 & J2000.0 & J2000. & &  \\
\hline
12175 & 247.81591 & 12.58238 & 18.72  & 0.83 & 35 & * \\
42241 & 247.73849 & 12.78898 & 18.72  & 0.82 & 36 & *  \\
41082 & 247.84564 & 12.74666 & 19.05  & 0.78 & 23 &  \\
42149 & 247.74718 & 12.79045 & 19.21  & 0.70 & 25 & * \\
41743 & 247.78386 & 12.80170 & 19.44  & 0.70 & 24 & *  \\
42795 & 247.68541 & 12.82996 & 19.51  & 0.67 & 23 & *  \\
40789 & 247.87404 & 12.74030 & 19.52  & 0.67 & 20 & *  \\
41460 & 247.80860 & 12.75741 & 19.60  & 0.69 & 21 & *  \\
42096 & 247.75261 & 12.82550 & 19.59  & 0.66 & 20 & *  \\
40993 & 247.85432 & 12.75811 & 19.73  & 0.67 & 20 & *  \\
42324 & 247.73111 & 12.76968 & 19.72  & 0.62 & 13 & *  \\
12729 & 247.78123 & 12.52606 & 19.84  & 0.67 & 12 & *  \\
40222 & 247.93108 & 12.78307 & 20.01  & 0.62 & 11 &  \\
42692 & 247.69607 & 12.75570 & 20.02  & 0.63 & 12 & \\
43688 & 247.59341 & 12.86022 & 20.04  & 0.61 &  8 &  \\
43428 & 247.61721 & 12.75078 & 20.09  & 0.60 & 11 &  \\
11239 & 247.87333 & 12.58958 & 20.11  & 0.61 &  9 &  \\
41912 & 247.76877 & 12.77069 & 20.15  & 0.64 &  8 &  \\
42008 & 247.76005 & 12.80071 & 20.21  & 0.61 &  9 &  \\
41371 & 247.81831 & 12.83070 & 20.23  & 0.63 & 10 &  \\
\hline
\end{tabular}
\begin{list}{}{}
\item[] Column 1 lists the RGB star ID \citep{2009A&A...506.1147A}. Columns 2 and 3 list the
  coordinates.  Column 4 lists the $V$ magnitude and column 5 lists the $(b-y)$ colour. Column 6 lists
  estimates of the S/N in the final spectra and column 7 indicates whether the star was analysed in this work,
  compare Sect. \ref{abund}.
\end{list}
\end{table}

\subsection{Observations}

Our spectroscopy was carried out using the multiobject spectrograph
Fibre Large Array Multi Element Spectrograph (FLAMES) at the Very
Large Telescope (VLT) on Paranal. The observations, 18 observing blocks of 60
minutes each made in service mode,
are summarised in Table \ref{table:3}.
Operated in Medusa fibre mode, this
instrument allows for the observation of up to 130 targets at the same
time \citep{2002Msngr.110....1P}. 23 fibres were dedicated to
observing blank sky. We used the GIRAFFE/HR13 grating, which provides
a nominal spectral resolution of $R \sim 20\, 000$ and a wavelength
coverage from $6100$ \AA\, to $6400$ \AA. We verified the spectral resolution
by measuring the full-width-half maximum of telluric emission lines in the combined
sky spectrum.

\begin{table}
\caption{Summary of the spectroscopic observations with FLAMES.}
\label{table:3}
\centering                     
\begin{tabular}{c c}     
\hline\hline          
Date & Exp. time [$min$] \\   
\hline
17 May 2009 & 180 \\  
20 May 2009 & 120 \\
22 May 2009 & 120 \\
23 May 2009 & 120 \\
24 May 2009 & 60 \\
25 May 2009 & 180 \\
26 May 2009 & 60 \\
13 June 2009 & 180 \\
18 June 2009 & 60 \\
\hline
Total Exp. Time & 1080 \\
\hline
\end{tabular}
\begin{list}{}{}
\item[] Column 1 lists the date of observation and column 2 the
  exposure time.
\end{list}
\end{table}

\subsection{Data reduction and measurement of equivalent widths}\label{reduction}

The FLAMES observations were reduced with the standard GIRAFFE
pipeline, version 2.8.1 \citep{2000SPIE.4008..467B}. This pipeline
provides bias subtraction, flat fielding, dark-current subtraction,
and accurate wavelength calibration from a ThAr lamp.

The 23 sky spectra were combined and subtracted from the object
spectra with the task SKYSUB in the SPECRED package in IRAF
\footnote{IRAF is distributed by the National Optical Astronomy
  Observatories, which are operated by the Association of Universities
  for Research in Astronomy, Inc., under cooperative agreement with
  the National Science Foundation.}.

Next, the object spectra from the individual frames were
Doppler-shifted to the heliocentric rest frame and median-combined
into the final one-dimensional spectrum. When combining the object
spectra we used an average sigma clipping algorithm, rejecting
measurements deviating by more than $3\, \sigma$, in order to remove
cosmic rays.

Finally, we normalised the spectra with the task CONTINUUM in the
ONEDSPEC package in IRAF. We used a Spline1 function of the 1st order.
We note that the nomalisation was not optimal over the entire wavelength range.
To accomodate for this, we set the continuum for each line individually when measuring
the equivalent widths, $W_{\lambda}$.

The$W_{\lambda}$ for the absorption lines
were measured by fitting a Gaussian profile
to each of the lines using the IRAF task
SPLOT. However, for some of the weak lines with low S/N it was better to determine the
$W_{\lambda}$ by integration of the pixel values using the "e" option in SPLOT.
The $W_{\lambda}$s are listed in Table \ref{table_ew}. \\

We were not able to identify any absorption lines in the continuum for
stars fainter than $V_0=19.80$. The S/N for the spectra for these stars are about 10.
Thus, 8 stars were discarded from the abundance analysis (compare Table \ref{table_obs}).
Additionally, we were not able to remove the sky emission for RGB star 41082 to a
satisfying level, and the S/N was lower than expected from the stars magnitude, indicating that something may have gone wrong with the
positioning of the fibre. Therefore, the spectrum for this star
was discarded also, leaving us with spectra for 11 usable RGB targets.

\begin{table*}
\caption{Equivalent width measurements.}
\label{table_ew}
\centering
\begin{tabular}{c c c c c c c c c c c c c c c}
\hline\hline
& & & RGB & 12175 & 42241 & 42149 & 41743 & 42795 & 40789 & 41460 & 42096 & 40993 & 42324 & 12729 \\
\hline
Ion & $\lambda$ & $\log gf$ & EP & $ W_{\lambda}$ & $ W_{\lambda}$ & $ W_{\lambda}$ & $ W_{\lambda}$ & $ W_{\lambda}$ & $ W_{\lambda}$ & $ W_{\lambda}$ & $ W_{\lambda}$ & $ W_{\lambda}$ & $ W_{\lambda}$ & $ W_{\lambda}$ \\
 & (\AA) & (dex) & eV & (m\AA) & (m\AA) & (m\AA) & (m\AA) & (m\AA) & (m\AA) & (m\AA) & (m\AA) & (m\AA) & (m\AA) & (m\AA) \\
\hline
Ca\,{\sc i}& 6122.22 & -0.386 & 1.886 & 46 & 89 & 44 & 66  & ...  & 24 & 49 & 89 & 52 & 57 & ...  \\
Ca\,{\sc i}& 6162.17 & -0.167 & 1.899  & 72 & 122 & 35 & 96 & ...  & 43 & ...  & 70 & ...  & 61 & 69 \\
Fe\,{\sc i}& 6137.69 & -1.403 & 2.588  & 41  & 148 & 48 & 65 & ...  & 43 & ...  & ...  & 66 & ...  & 86 \\
Fe\,{\sc i}& 6151.62 & -3.299  & 2.176 & ...   & 62  & ...  & ...   & ...  & ...  & ...  & ...  & ...  & ...  & ... \\
Fe\,{\sc i}& 6173.34 & -2.880 & 2.223  & ...   & 96  & ...  & 52  & ...  & ...  & ...  & ...  & 65 & ...  & ...	 \\
Fe\,{\sc i}& 6180.20 & -2.586 & 2.727  & ...   & 38  & ...  & 13  & ...  & ...  & ...  & ...  & 30 & ...  & ...	 \\
Fe\,{\sc i}& 6200.31 & -2.437 & 2.608  & ...   & 75  & ...  &       & ...  & 14 & ...  & ...  & ...  & ...  & ...	 \\
Fe\,{\sc i}& 6213.43 & -2.482 & 2.223  & ...   & 92  & 25 & 35  & ...  & ...  & ...  & ...  & 49 & ...  & 65 \\
Fe\,{\sc i}& 6219.28 & -2.433 & 2.198  & ...   & 111 & ...  & 58 & ...  & ...  & ... & ...  & 40 & ...  & ...	 \\
Fe\,{\sc i}& 6232.64 & -1.223 & 3.654  & ...   & 59  & ...  & ...   & ...  & ...  & ...  & ...  & ...  & ...  & ...	 \\
Fe\,{\sc i}& 6246.32 & -0.733 & 3.602  & ...   & 87  & ...  & ...   & ...  & ...  & ...  & ...  & ...  & ...  & ...	 \\
Fe\,{\sc i}& 6252.56 & -1.687 & 2.404  & 50  & 149 & ...  & 81  & ...  & ...  & ...  & 71 & ...  & ...  & 98 \\
Fe\,{\sc i}& 6265.13 & -2.550 & 2.176  & ...   & 124 & ...  & 83  & ...  & ...  & ...  & 43 & ...  & ...  & ...	 \\
Fe\,{\sc i}& 6270.23 & -2.464 & 2.858  & ...   & 42  & ...  & ...   & ...  & ...  & ...  & ...  & ...  & ...  & ...	 \\
Fe\,{\sc i}& 6301.50 & -0.718 & 3.654 & ...   & 119 & ...  & 42  & ...  & ...  & ...  & ...  & ...  & ... & ...	 \\
Fe\,{\sc i}& 6302.49 & -0.973 & 3.686  & ...   & 57  & ...  & ...   & ...  & ...  & ...  & ...  & ...  & ...  & ...	 \\
Fe\,{\sc i}& 6322.69 & -2.426 & 2.588 & ...   & 97  & ...  & 40 & ...  & ...  & ...  & ...  & ...  & ...  & ...	 \\
Fe\,{\sc i}& 6335.33 & -2.177 & 2.198 & 36  & 121 & ...  & 91  & 27 & ...  & ...  & 58 & 51 & ...  & 50 \\
Fe\,{\sc i}& 6336.82 & -0.856 & 3.686 & ...   & 83  & ...  & ...   & ...  & ...  & ...  & ...  & 33 & ...  & ...	 \\
Fe\,{\sc i}& 6355.03 & -2.350 & 2.845 & ...   & 55  & ...  & ...   & ...  & ...  & ...  & ...  & ...  & ...  & ...	 \\
Fe\,{\sc i}& 6393.60 & -1.432 & 2.433 & 69  & 138 & ...  & 109 & 35 & 39 & ...  & 65 & 68 & ...  & 85 \\
Fe\,{\sc ii} & 6149.26 & -2.841 & 3.899 & ...   & 30  & ...  & ...   & ...  & ...  & ...  & ...  & ...  & ...  & ...  \\
Fe\,{\sc ii} & 6247.56 & -2.435 & 3.892 & ...   & 40  & ...  & ...   & ...  & ...  & ...  & ...  & ...  & ...  & ...	 \\
\hline
\end{tabular}
\begin{list}{}{}
\item[] {\bf}
\end{list}
\end{table*}

\section{Stellar parameters} \label{stepar}

The effective temperature
($T_{eff}$) is often determined by
requiring that the abundances derived from individual Fe lines are independent of the
excitation potential for the lines. This was not an option for us due to the small number of Fe {\sc i}
lines for each star.  Instead, we calculated
$T_{eff}$ from Str\"omgren photometry
using the calibration in \citet{1999A&AS..140..261A}.
The photometry is from \citet{2009A&A...506.1147A} and has been corrected
for interstellar extinction using the dust maps by \citep{1998ApJ...500..525S}. These maps give a reddening of $E(B-V)=0.062$.
We estimated the errors in $T_{eff}$ using the uncertainties
for the Str\"omgren photometry. Since we are using deep photometry, and are only using
stars at the brighter end of the luminosity function, the errors are essentially the same for the stars
in the sample. We find a typical error
of about $100$ K for all stars.

\begin{figure}
\resizebox{\hsize}{!}{\includegraphics{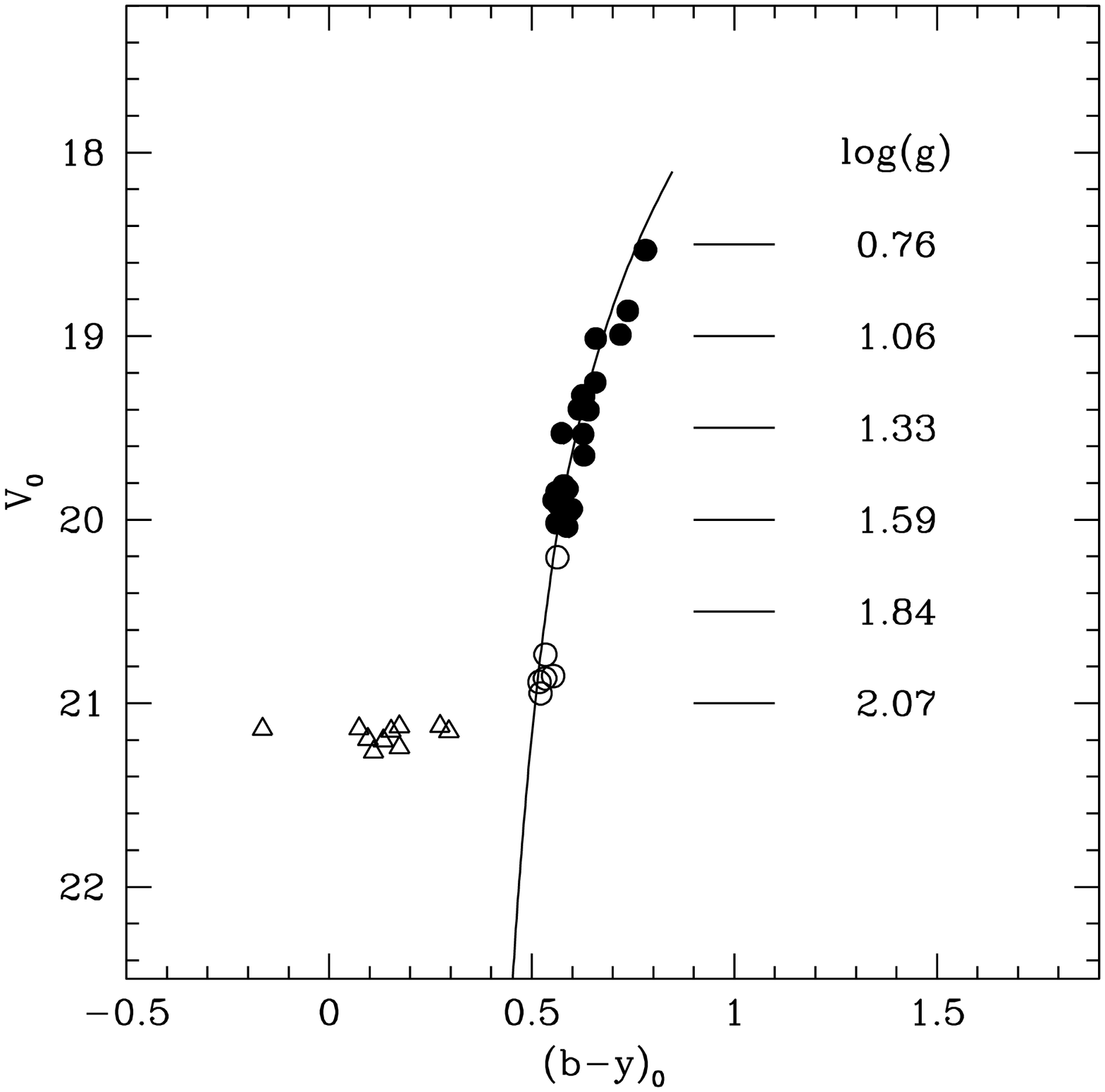}} 
\caption{Colour-magnitude diagram for the Hercules dSph galaxy \citep{2009A&A...506.1147A}.
  $\bullet$ are RGB stars selected for this study. $\circ$ indicate
  RGB stars too faint for this study and open triangles are
  horizontal-branch stars. The solid line indicates the isochrone by
  \citet{2006ApJS..162..375V} with colour transformations by
  \citet{2004AJ....127.1227C}. $\log g$ values for different magnitudes as
  indicated. Note that there are two stars at $V_0 \sim 18.5$ superimposed on each other.}
\label{iso}
\end{figure}

Surface gravities, $\log g$, were estimated using an isochrone by
\citet{2006ApJS..162..375V} with ${\rm [Fe/H]=-2.31}$ (most metal-poor isochrone available), an age of 12 Gyr, 
colour transformations by
\citet{2004AJ....127.1227C}, and no $\alpha$-enhancement. Figure \ref{iso} shows the
colour-magnitude diagram for the Hercules dSph galaxy with $\log g$ values
indicated. The isochrone was shifted using the
distance modulus derived in \citet{2009A&A...506.1147A}, $(m-M)=20.85 \pm 0.11$.

To estimate how sensitive our value of $\log g$ is to the choice of the age for the isochrone,
we repeated the above derivation for isochrones with an age of 8 and 18 Gyr,
and ${\rm [Fe/H]=-2.31}$. We find that the estimated value of $\log g$
deviated by a maximum of $\sim 0.1$ dex from the initial $\log g$ when the age was changed.
Additionally, for comparison with an isochrone based on a different stellar evolutionary model, we compared with values of $\log g$ derived using the Darthmouth isochrones \citep{2008ApJS..178...89D} with colour transformations by \citet{2004AJ....127.1227C}, and similar age and metallicity as for the isochrone by
\citet{2006ApJS..162..375V}. We find that the values of $\log g$ estimated using the two sets of
isochrones differ by about 0.1 dex.

Finally, we estimated the contribution to the error in $\log g$ from the uncertainty in the distance modulus and magnitude using $10^6$ Monte Carlo realisations of the distance modulus and magnitude drawn from within the individual error bars on each parameter.
We find that the values of $\log g$ deviated by $\sim 0.1$ dex from the initial $\log g$.
Based on these three error estimates, we define an upper limit to the error in $\log g$ of 0.35 dex to make
sure that the error is not under-estimated.

In Sect. \ref{abund} we investigate how different values
of $\log g$ affect the abundance analysis.

\begin{figure}
\resizebox{\hsize}{!}{\includegraphics{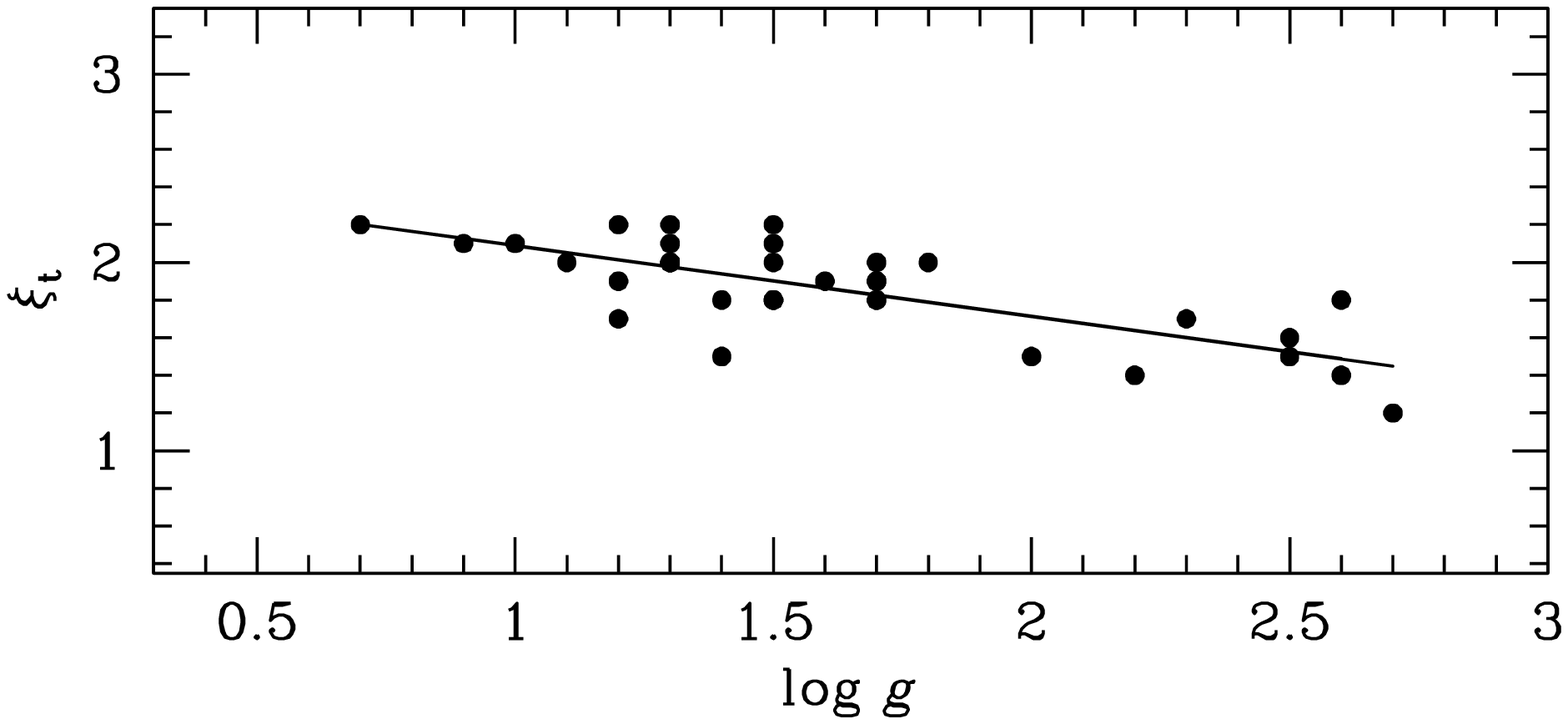}} 
\caption{$\xi_t$ vs. $\log g$ for metal-poor giant stars
  from \citet{2010A&A...509A..88A}.  The solid line indicates a
  least-square fit to the data.}
\label{xi}
\end{figure}

We estimated the microturbulence, $\xi_t$, using the $\xi_t$ and $\log
g$ for metal-poor halo stars from
\citet{2010A&A...509A..88A}. These stars have about the same
metallicity and $\log g$ as our Hercules RGB stars.  A least-square fit to their
data, in $\xi_t$ vs. $\log g$ space (Fig. \ref{xi}), of 35 giant stars yields
\begin{equation} \label{xi_eq}
\xi_t=-0.38(\pm 0.06) \cdot \log g +2.47(\pm 0.1).
\end{equation}

We estimated the errors in $\xi_t$ using the uncertainties
for the least-square fit (Eq. \ref{xi_eq}) and an uncertainty
in $\log g$ of 0.3 dex. We find a typical error in $\xi_t$
of $\sim 0.2$ ${\rm km \, s^{-1} }$.

The final stellar parameters used in the abundance analysis are summarised in Table \ref{table:1}.

\begin{table*}
\caption{Photometry and model parameters used in the abundance analysis of the stars.}
\label{table:1}
\centering
\begin{tabular}{c c c c c c c c c}
\hline\hline
ID & Other ID &  $V_0$ & $(b-y) _0$ & [Fe/H] & $T_{eff}$ & $ \log g$ & $\xi_t$ & S/N \\   
 & &   &  & dex & K & dex & ${\rm km \, s^{-1} }$ \\   
\hline
12175 & ... &  18.53 & 0.78 & -3.17 & 4370 & 0.78  & 2.17 & 35  \\
42241 & Her-2 &  18.53 & 0.78 & -2.03 & 4270 & 0.78  & 2.17 & 36  \\
41082 & Her-3 & 18.86 & 0.74 & ... & 4340 & 0.97  & 2.10 & 23  \\
42149 & ... &  19.01 & 0.66 & -2.95 & 4540 & 1.06  & 2.07 & 25  \\
41743 & ... &  19.25 & 0.66 & -2.42 & 4520 & 1.19  & 2.02 & 24  \\
42795 & ... &  19.32 & 0.63 & -3.17 & 4620 & 1.22  & 2.01 & 23  \\
40789 & ... &  19.33 & 0.63 & -2.88 & 4600 & 1.24  & 2.00 & 20  \\
41460 & ... &  19.40 & 0.64 & -3.10 & 4590 & 1.27  & 1.99 & 21  \\
42096 & ... &  19.40 & 0.62 & -2.60 & 4620 & 1.27  & 1.99 & 20  \\
40993 & ... &  19.53 & 0.63 & -2.38 & 4600 & 1.33  & 1.97 & 20  \\
42324 & ... &  19.53 & 0.57 & -2.70 & 4740 & 1.33  & 1.97 & 13  \\
12729 & ... &  19.65 & 0.63 & -2.35 & 4640 & 1.40  & 1.94 & 12  \\
\hline
\end{tabular}
\begin{list}{}{}
\item[] Column 1 lists the RGB star ID \citep{2009A&A...506.1147A}. Column 2 lists the ID
from \citet{2008ApJ...688L..13K}.
Column 3 lists the $V_0$ magnitude and column 4 lists the $(b-y)_0$ colour. Column 5 lists the
  metallicity as determined in Sect. \ref{iron}.
 Column 6 to 8 list the stellar
  parameters as determined in Sect. \ref{stepar}. Column 9 lists an
  estimate of the S/N.
\end{list}
\end{table*}

\section{Abundance analysis} \label{abund}
Model atmospheres were calculated for the programme stars with the code {\sc marcs} according to the 
procedures described in \citet{2008A&A...486..951G} and using the fundamental parameters in Table \ref{table:1}.
Next a line list was compiled in the wavelength region 6120 -- 6400 \AA\, with spectral lines
from neutral and singly ionised atoms from the VALD database \citep{1995A&AS..112..525P,1997BaltA...6..244R,1999A&AS..138..119K,2000BaltA...9..590K}.
Equivalent widths or synthetic spectra were then computed from radiative transfer calculations in
spherical geometry in the model atmospheres using the Eqwi/Bsyn codes that share many
subroutines and data files with {\sc marcs} making the analysis largely self-consistent.

For stars with at least two lines measurable, we adopt the mean of the abundances
derived from the individual $W_{\lambda}$ for each element as the final elemental abundances. For stars more difficult, in terms of identifying absorption lines, the final elemental abundances are determined using a $\chi^2$-test (see Sect. \ref{iron_difficult})

\subsection{Elemental abundance errors}\label{err}

For elements with more than four lines measured,
the random errors in the elemental abundance ratios were calculated as
\begin{equation}
\epsilon_{{\rm rand, [X/H]}}=\frac{\sigma_X}{\sqrt{N}}
\end{equation}
where $X$ is the element, $\sigma$ is the standard deviation of the abundances derived from 
the individual $W_{\lambda}$, and $N$ the number of lines for that
element.
For elements with two to four lines measured, the uncertainty in the measurement of $W_{\lambda}$, $\epsilon_{W_{\lambda}}$, was estimated using the relation in \citet{1988IAUS..132..345C}. The random errors in the elemental abundances were then estimated using $10^5$ Monte Carlo realisations of ${W_{\lambda}}$, drawn from within the probability distribution of ${W_{\lambda}}$ given $\epsilon_{W_{\lambda}}$. For each value of ${W_{\lambda}}$, we recalculate an elemental abundance using the relation $\log(A) \propto \log(W_{\lambda})$ where $A$ is the elemental abundance. We note that the probability distribution of $\log(A)$ is asymmetric. Thus, we adopt the standard deviation based on the sextiles (which is equivalent to 1$\sigma$ in the case of a Gaussian distribution) as our final random error.
For elements with less than two lines measured we performed a $\chi^2$-test
between the stellar spectrum and a grid of
synthetic spectra, to estimate the random errors in the elemental abundances (see
Sect. \ref{iron} and \ref{calcium}). Note that none of the Ca abundances are estimated using more than two lines. Thus, the errors in [Ca/H] are derived using either a $\chi^2$-test or by propagating $\epsilon_{W_{\lambda}}$ as derived using the relation in \citet{1988IAUS..132..345C}.

The systematic errors, $\epsilon_{{\rm sys, [X/H]}}$,  were estimated from the errors in
the stellar parameters (Sect. \ref{stepar}) as follows: two stars were selected randomly, RGB star 40789 and 42241. 
For these two stars, we study the final
elemental abundances for several model atmospheres. The model atmospheres were
chosen so that we had two values of $\log g$, separated by 0.5 dex, three values of $T_{eff}$, separated by 100 K, and three values of $\xi_t$, separated by 0.2 ${\rm km \, s^{-1} }$.
The separation between the $T_{eff}$ and $\xi_t$ values corresponds to the estimated errors in the parameters
(see Sect. \ref{stepar}). The centre value for $T_{eff}$ and $\xi_t$ corresponds to the
values as determined in Sect. \ref{stepar}. Since the error in $\log g$ was more difficult to determine (see Sect. \ref{stepar}), we chose a separation in $\log g$ of 0.5 dex to make sure that we got an upper limit of the contribution from this stellar parameter. The elemental abundances varies with less than 0.05 dex when the value of $\log g$ is separated by 0.5 dex. However, we note that this is based on Fe\,{\sc i} lines. Fe\,{\sc ii} lines are more sensitive to changes in $\log g$.

Thus, we have 18 model atmospheres for which we determine the final elemental abundances
of iron and calcium.
The standard deviation of the 18
final elemental abundances, for iron and calcium,
is then adopted as $\epsilon_{{\rm sys, [X/H]}}$.
We find a typical $\epsilon_{{\rm sys, [X/H]}}$
of $\sim 0.12$ dex.

The total errors in the elemental abundance ratios were calculated as
\begin{equation}
\epsilon_{{\rm [X/H]}}= \sqrt{\epsilon_{{\rm rand, [X/H]}}^2+\epsilon_{{\rm sys, [X/H]}}^2}
\end{equation}
The final total errors are summarised in Table \ref{table:2}.

\subsection{Iron} \label{iron}

The mean [Fe/H] is determined on the scale where $\log \epsilon_{\rm
  H}=12.00$. The solar iron abundance of 7.45 is adopted from
\citet{2007SSRv..130..105G}.

Due to the variation in the S/N, and the number of measurable lines in
the spectra, we analyse these stars individually or as groups with spectra
of similar quality (Sections \ref{iron_high}, \ref{iron_low} and \ref{iron_difficult}).  The result from the analysis is summarised
in Table \ref{table:2}.

\subsubsection{Highest S/N spectra} \label{iron_high}

\begin{figure*}
\resizebox{\hsize}{!}{\includegraphics{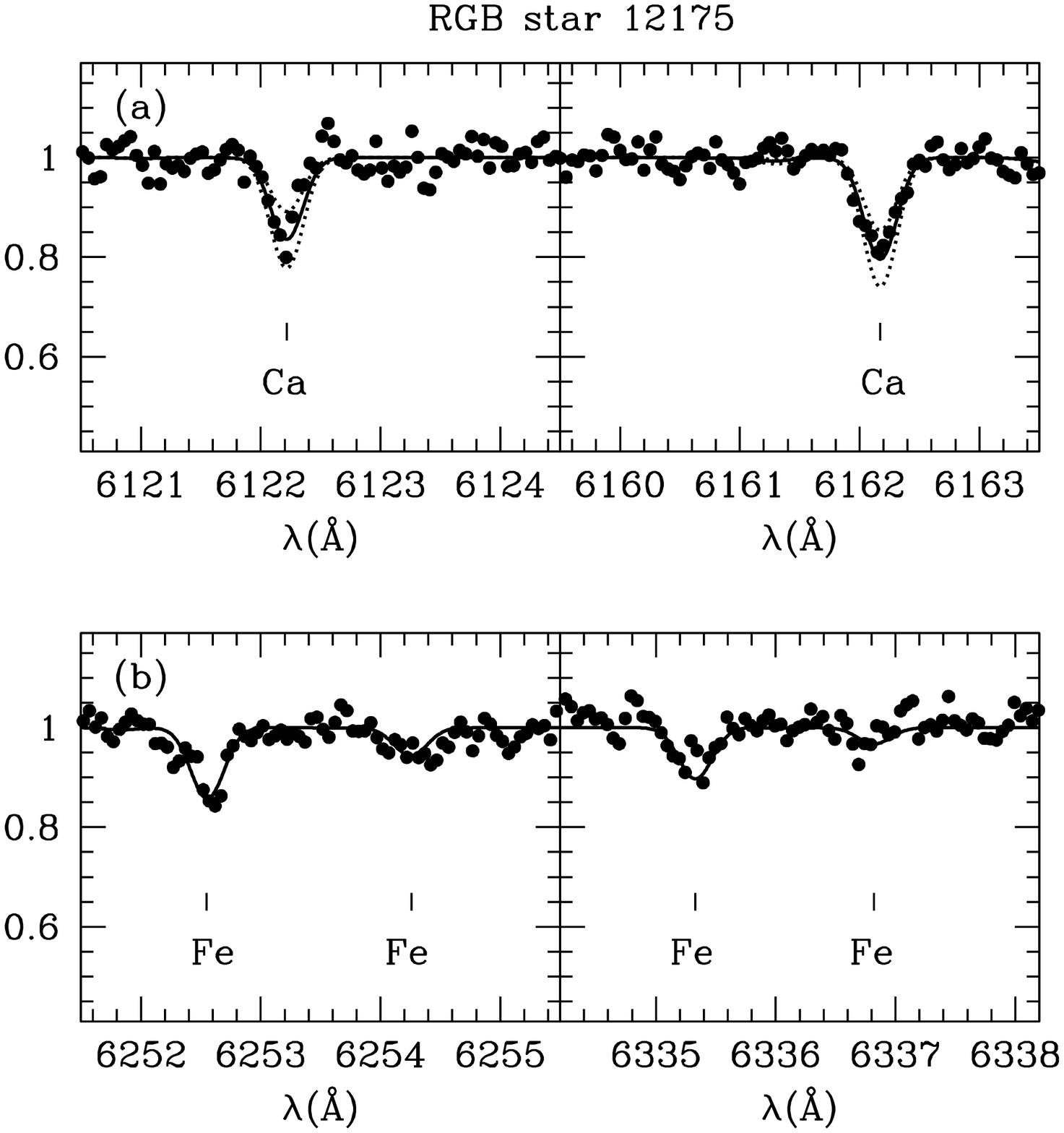}\includegraphics{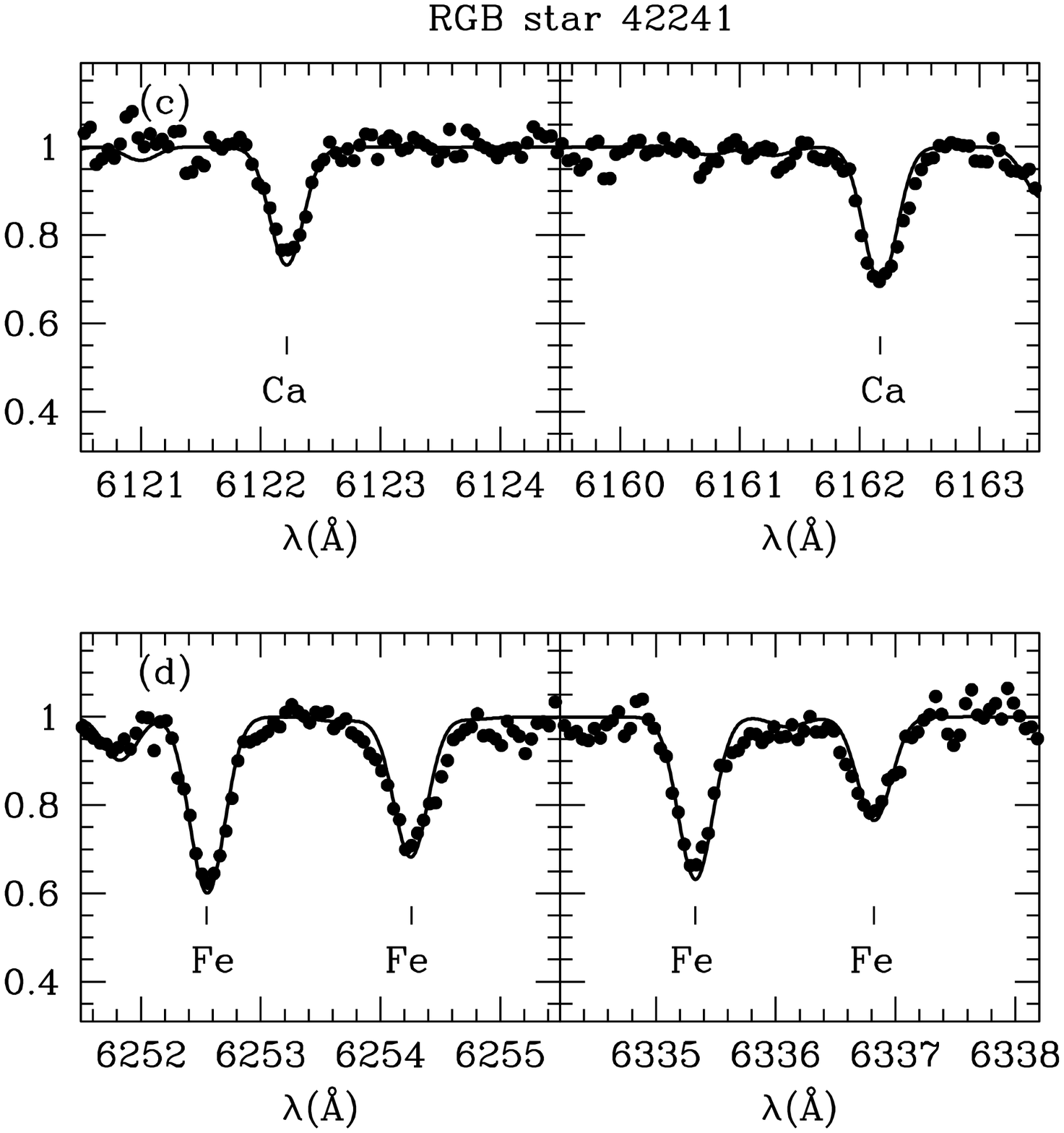}} 
\caption{{\bf Left panels:} Portions of stellar spectra around four Fe\,{\sc i} {\bf (b)} and two Ca\,{\sc i} {\bf (a)} lines for RGB star 12175.  
 $\bullet$ indicate
  the observed spectrum. The solid line indicates a synthetic spectrum
  with ${\rm [Fe/H]=-3.1}$ and ${\rm [Ca/H]=-2.8}$. The dotted lines in {\bf (a)} indicate synthetic spectra with ${\rm [Ca/H]\pm 0.3}$ relative to the solid-line-synthetic spectrum.
  {\bf Right panels:} Portions of stellar spectra around four Fe\,{\sc i} {\bf (d)} and two Ca\,{\sc i} {\bf (c)} lines
  for star 42241. $\bullet$ indicate
  the observed spectrum. The solid line indicates a synthetic spectrum
  with ${\rm [Fe/H]=-2.0}$ and ${\rm [Ca/H]=-2.6}$.}
\label{rgb_11}
\end{figure*}

RGB star 12175, with $V_0=18.5$, is one of the two brightest RGB stars
discovered in the Hercules dSph galaxy. However, due to its low
metallicity, only 4 Fe\,{\sc i} lines were distinguishable from the continuum
in the spectral range covered by our observation.  These four iron
lines give [Fe/H] $=-3.17\pm 0.14$. Figure \ref{rgb_11}b shows the
spectrum of 12175 around two of the four Fe\,{\sc i} lines. Close to these 
two lines there are two additional Fe\,{\sc i} lines that we could not measure
quantitatively, but that we were able to identify with the help of a synthetic
spectrum.  The synthetic spectrum shown in
Fig. \ref{rgb_11}b supports the result that this is a very metal-poor
star with [Fe/H]$=-3.2$.

RGB star 42241 has about the same magnitude and S/N as
12175. However, due to its higher iron abundance, 
about four times as many lines were measurable in this
spectrum (compare Table \ref{table_ew}). We find an [Fe/H] of
$-2.03\pm 0.14$ dex. Additionally, for this star, we were able to
measure two Fe\,{\sc ii} lines. These lines give an iron abundance of
$-1.40\pm 0.20$ dex. Thus, the [Fe/H] as derived from Fe\,{\sc i}
lines do not agree within the error bars with [Fe/H] as derived from
Fe\,{\sc ii} lines. This discrepancy in the determination of the iron abundance could partially be caused by over-ionisation in Fe\,{\sc i}. \citet{2001AJ....122.1438I} argue that over-ionisation could cause an under-estimate of about 0.1 dex for RGB stars if Fe\,{\sc i} lines are used.

Figure \ref{rgb_11}d shows the stellar spectrum of
42241 around four of the measured Fe\,{\sc i} lines. As can
be seen, [Fe/H] derived from the $W_{\lambda}$ agrees well with a
synthetic spectrum with an iron abundance close to the --2 dex value derived from the $W_{\lambda}$s.

\subsubsection{Low S/N spectra} \label{iron_low}

RGB stars 42149, 41743, 42795, 40789, 42096, 40993 and 12729 have a
lower S/N than 12175 and 42241. However, at
least two Fe\,{\sc i} lines were measurable for each of the stars. 

Since
the S/N is much lower for these stars, we did the following test to
ensure that the [Fe/H] derived from the equivalent widths are
reasonable.  For each of the stars, we generated a set of synthetic
spectra with five different [Fe/H] values, separated by 0.2 dex, centred on
the [Fe/H] derived from the equivalent widths.  A plot of the
stellar spectrum, with the synthetic spectra over-plotted, enabled us
to verify that the [Fe/H] derived from the $W_{\lambda}$s is a good estimate
of the iron abundance of the star. We found that none of the stellar spectra deviated
significantly from a synthetic spectrum with a similar [Fe/H]
abundance. 

\begin{figure}
\resizebox{\hsize}{!}{\includegraphics{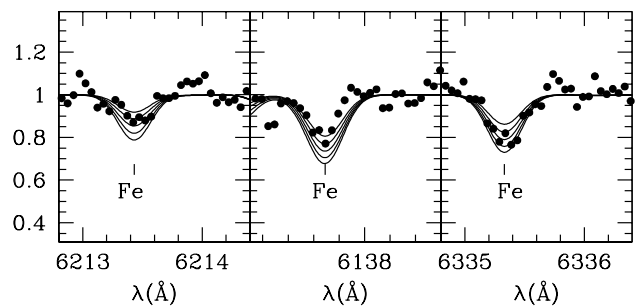}} 
\caption{Portions of stellar spectra around three Fe\,{\sc i} lines
  for RGB star 41743.  $\bullet$ indicate the observed
  spectra. The solid lines indicate synthetic spectra with [Fe/H]
  from $-2.85$ to $-2.05$ dex, top to bottom, separated by 0.2 dex each.}
\label{rgb_20}
\end{figure}

Figure \ref{rgb_20} shows an example, for 41743. As can be seen, an [Fe/H] 
of $\sim -2.4$ dex is a reasonable estimate of the iron abundance for this star.

\subsubsection{Difficult spectra} \label{iron_difficult}

RGB stars 41460 and 42324 have low S/N (21 and 13, respectively) and are very metal-poor. Thus,
it was difficult to identify the Fe\,{\sc i} absorption lines in the
spectra. However, we did see faint absorption signatures but the low S/N made it virtually impossible
to measure the lines. Instead, we performed a $\chi^2$-test between the observed spectrum
and a grid of synthetic spectra with 17 different [Fe/H] values, separated by 0.05 dex. Each synthetic spectrum yields a $\chi^2$ value, and the best fit is found when $\chi^2$ is minimised ($\chi^2_{min}$). We used a width of 3 $\sigma$, which covers about 99.7 per cent of the absorption feature, for each iron line in the line list (see Table \ref{table_ew}). We investigated the sensitivity of the $\chi^2$-test region by varying it between 2 $\sigma$ and 4 $\sigma$ and found that it had negligible impact on the result. The continuum for each line was adjusted, as the average of the signal on each side of the absorption feature over ~0.6 \AA, to accommodate for the local deviations from the continuum normalisation in Sect. \ref{reduction}. 
The error for each pixel in the observed spectrum was approximated by the variance in the spectrum.
The distribution enclosed by $\chi^2_{min}+1$ corresponds to 1 $\sigma$ for a normal distribution \citep{1992nrfa.book.....P}. We used this as the measurement error.

\subsection{Calcium} \label{calcium}

The mean [Ca/H] is determined on the scale where $\log \epsilon_{\rm
  H}=12.00$ The solar calcium abundance of 6.34 is adopted from
\citet{2009ARA&A..47..481A}.  There are two Ca\,{\sc i} lines, at 6122.22 \AA \,
and 6162.17 \AA\, in the wavelength region covered by our spectra,
that were possible to measure in the
majority of the stars.  The results from the
analysis are summarised in Table \ref{table:2}.
Individual stars are, in similar manner as done for Fe, discussed in the following sections.

\subsubsection{Highest S/N spectra}

For RGB star 12175 both Ca\,{\sc i} lines were measured and they give
[Ca/H]$=-2.9 \pm 0.1$.  Figure \ref{rgb_11}a shows the stellar
spectrum around the Ca\,{\sc i} lines. As can be seen,
[Ca/H] as measured from the $W_{\lambda}$ agree well with a synthetic
spectrum with similar [Ca/H]. Additionally, Fig. \ref{rgb_11}a shows an example
of two synthetic spectra with [Ca/H] $\pm 0.3$. 

For RGB star 42241 we find a [Ca/H] of
$-2.5\pm 0.2$. This is significantly higher than [Ca/H] for RGB star
12175. Figure \ref{rgb_11}c shows the stellar spectrum of 42241 around
the two Ca\,{\sc i} lines. As can be seen, [Ca/H] as derived from the
$W_{\lambda}$ agree well with a synthetic spectrum with a similar
abundance.

\subsubsection{Low S/N spectra}

RGB stars 42149, 41743, 40789, 42096 and 42324 have a lower S/N
than RGB stars 12175 and 42241. However, both of the Ca\,{\sc i} lines were
measurable in all four stars.  

Since the S/N is lower
we repeated the same test done for the
iron abundance analysis (compare Sect. \ref{iron_low}),
generating a grid of synthetic spectra for each star, to see if the
[Ca/H] as derived from the $W_{\lambda}$ were reasonable.  We found
that the synthetic spectrum of RGB star 40789, in comparison with the
observed spectrum, indicates that the [Ca/H] determined from the
measurements of the $W_{\lambda}$ was slightly, about 0.1 dex, over-estimated.  
Thus, we estimated the Ca\,{\sc i} abundance for RGB star 40789 using the same method as for the spectra identified as difficult for the measurement of the Fe\,{\sc i} lines (see
Sect. \ref{iron_difficult}). For all other stars in this category, the abundances from the measured  $W_{\lambda}$ and those from the $\chi^2$-comparison of synthetic spectra showed good agreement.

\subsubsection{Difficult spectra} \label{calcium_difficult}

For RGB star 42795, 41460, and 40993 only one or none of the Ca\,{\sc i} lines
were measurable.  However we did see a general decrease in the
continuum at the wavelengths for the Ca\,{\sc i} lines indicating the presence
of Ca in the atmospheres of these metal-poor stars. Thus, we estimated
the Ca\,{\sc i} abundance using the same method as for the spectra identified
as difficult for the measurement of the Fe\,{\sc i} lines (see
Sect. \ref{iron_difficult}).  The results from the
analysis are summarised in Table \ref{table:2}.

We were not able to identify any Ca\,{\sc i} absorption
features for RGB star 12729.  Thus, [Ca/H] remains unknown for this
star.

\begin{table}
\caption{Derived elemental abundances for the RGB stars in the Hercules dSph galaxy.}
\label{table:2}
\centering
\begin{tabular}{c c c c c r}
\hline\hline
Star & [Fe/H] & $N$ & [Ca/H] & $N$ & [Ca/Fe] \\
\hline
12175 & $-3.17 \pm 0.14$ & 4 & $-2.89\pm 0.15$ & 2 & $0.28 \pm 0.21$ \\
42241 & $-2.03 \pm 0.14$ & 20 & $-2.54\pm 0.15$ & 2 & $-0.51\pm 0.21$ \\
42149 & $-2.95 \pm 0.15$ & 2 & $-3.08 \pm 0.16$ & 2 & $-0.13\pm 0.22$  \\
41743 & $-2.42 \pm 0.15$ & 11 & $-2.51 \pm 0.16 $ & 2 & $-0.09\pm 0.22$  \\
42795 & $-3.17 \pm 0.15$ & 2 & $-3.11\pm 0.17$ & $\chi^2$ & $0.06\pm 0.23$ \\
40789 & $-2.88 \pm 0.17$ & 3 & $-3.06 \pm 0.16$ & $\chi^2$ & $-0.18\pm 0.23$ \\
41460 & $-3.10 \pm 0.16$ & $\chi^2$ & $-2.78 \pm 0.15$ & $\chi^2$ & $0.32 \pm 0.22$ \\
42096 & $-2.60 \pm 0.17$ & 4 & $-2.40 \pm 0.18$ & 2 & $0.20\pm 0.25$ \\
40993 & $-2.38 \pm 0.19$ & 8 & $-2.68\pm 0.15$ & $\chi^2$ & $-0.3 \pm 0.24 $ \\
42324 & $-2.70 \pm 0.14$ & $\chi^2$ & $-2.60\pm 0.28$ & 2 & $ 0.10 \pm 0.31$  \\
12729 & $-2.35 \pm 0.17$ & 5 & ... & ...  \\
\hline
\end{tabular}
\begin{list}{}{}
\item[] Column 1 lists the RGB star ID. Column 2 and 4 list the [Fe/H]
  and [Ca/H], respectively, with total errors in the abundances as indicated. N indicates the number of lines measured
  for the determination of [Fe/H] and [Ca/H], as indicated. $\chi^2$
  indicates that the corresponding abundance was determined through a
  $\chi2$-test using a grid of synthetic spectra (see
  Sect. \ref{iron_difficult} and \ref{calcium_difficult}). Column 6 lists [Ca/Fe].
\end{list}
\end{table}

\section{A comparison with abundances determined in other studies} \label{comparison}

\subsection{A comparison with a high S/N RGB star} \label{comparison_high}

 \citet{2009A&A...503..545L} obtained high S/N spectroscopy of several bright RGB stars in the Milky Way. They used the same instrument and grating (GIRAFFE/HR13) as in this study. Through private communication they provided us with a spectrum of one of their bright targets, star 17691, that has a S/N of about 300. We measured the $W_{\lambda}$ for the lines in Table \ref{table_ew} and performed an abundance analysis for this star as described in Sect. \ref{reduction} and \ref{abund}. The stellar parameters was adopted from \citet{2009A&A...503..545L}.  We find an Fe abundance that is 0.01 dex more metal poor, and a Ca abundance 0.04 dex lower than given in \citet{2009A&A...503..545L}. Thus,  our determinations of the abundances of Ca and Fe in star 17691 are in agreement with \citet{2009A&A...503..545L}. Additionally, we find that none of the elemental abundances as derived from individual measurements of the $W_{\lambda}$ deviate significantly. This suggests that the effect of atomic parameters should not contribute to our elemental abundance errors.

\subsection{A comparison with earlier spectroscopic results} \label{comparison_spec}

\citet{2008ApJ...688L..13K} obtained high resolution spectroscopy ($R
\sim 20000$), with similar S/N and resolution as in this study, of two stars in the Hercules dSph galaxy, Her-2 and
Her-3. These stars correspond to our RGB stars 42241 and 41082.  However,
RGB star 41082 was discarded from our sample (see Sect. \ref{abund}).  
\citet{2008ApJ...688L..13K} find ${\rm [Fe/H]=-2.02\pm 0.20}$ and 
${\rm [Ca/Fe]=-0.13\pm 0.05}$ for RGB star 42241. Note, however, that \citet{2008ApJ...688L..13K}
measured $W_{\lambda}$ values of lines over a broader wavelength range from
5500--8900\,\AA.
Our estimates of [Fe/H]
are in very good agreement, but [Ca/Fe] as derived by \citet{2008ApJ...688L..13K}
is 0.4 dex higher.

\begin{table}
\caption{Equivalent width measurements for RGB star 42241/Her-2.}
\label{table_compare}
\centering
\begin{tabular}{c c c c c}
\hline\hline
Ion & $\lambda$ & $W_{\lambda, A}$ & $W_{\lambda,K}$ & $\frac{W_{\lambda,A}}{W_{\lambda,K}}$ \\   
 & (\AA) & (m\AA) & (m\AA) &  \\   
\hline
Ca\,{\sc i}& 6122.22 & 89 & 124 & 0.72 \\
Ca\,{\sc i}& 6162.17 & 122 & 153 & 0.79 \\
Fe\,{\sc i}& 6137.69 & 148 & 141 & 1.05 \\
Fe\,{\sc i}& 6151.62 & 62 & 49 & 1.26 \\
Fe\,{\sc i}& 6173.34 & 96 & 98 & 0.98 \\
Fe\,{\sc i}& 6180.20 & 38 & 40 & 0.94 \\
Fe\,{\sc i}& 6200.31 & 75 & 40 & 1.87 \\
Fe\,{\sc i}& 6213.43 & 92 & 111 & 0.83 \\
\hline
Blend & 6128.96 & 35 & 33 & 1.06 \\
Blend & $\sim$6136.6 & 239 & 241 & 0.99 \\
Blend & 6163.54 & 36 & 33 & 1.10 \\
Blend & $\sim$6191.5 & 265 & 262 & 1.01 \\
\hline
\end{tabular}
\begin{list}{}{}
\item[] Column 1 and 2 list the Ion and wavelength, respectively. Column 3 and 4 list
$W_{\lambda}$ as measured from our observed spectrum and the spectrum obtained by
\citet{2008ApJ...688L..13K}, respectively. Column 5 lists the ratio between the measurements.
\end{list}
\end{table}

\begin{figure*}
   \sidecaption
   \centering
   \includegraphics[width=14cm]{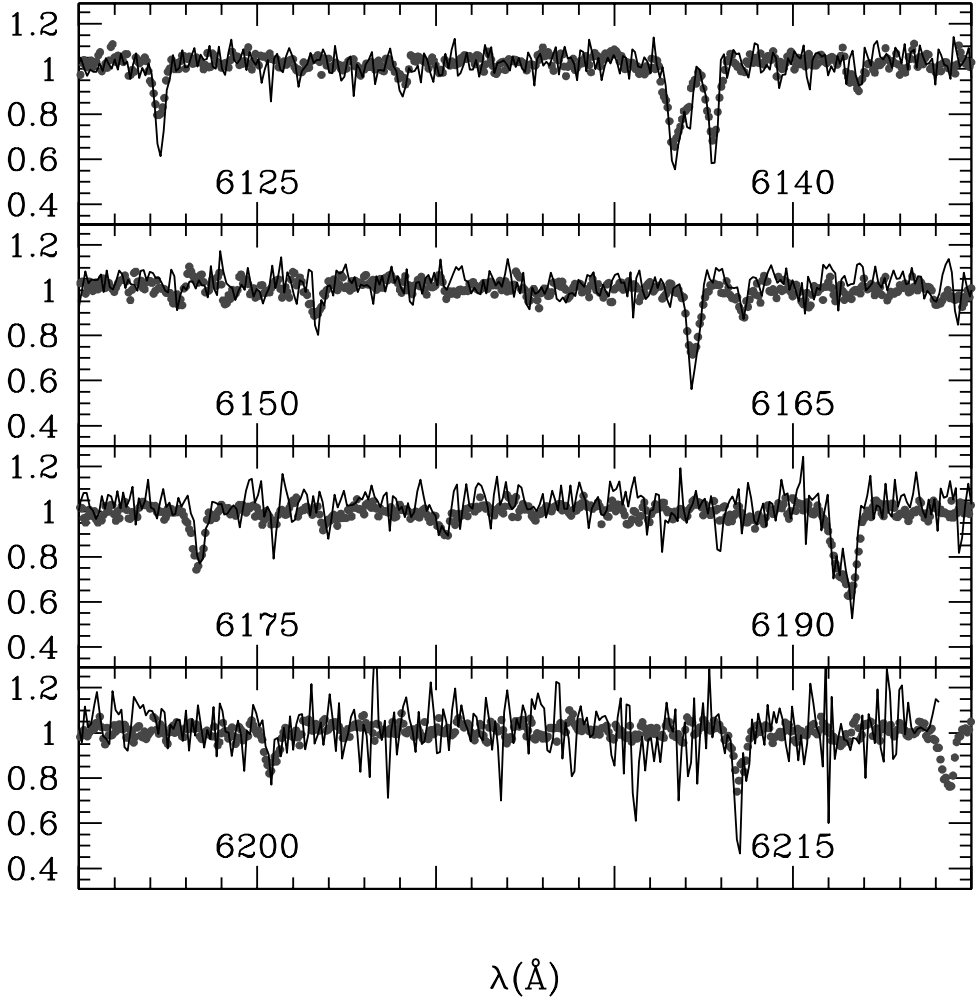}
      \caption{A comparison between our spectrum and the spectrum from
  \citet{2008ApJ...688L..13K} , in the region where they overlap, for
  RGB star 42241. $\bullet$ indicate our stellar spectrum. The solid
  line indicates the spectrum from \citet{2008ApJ...688L..13K}.
              }
         \label{comp}
   \end{figure*}

Figure \ref{comp} shows our spectrum and the spectrum from
\citet{2008ApJ...688L..13K} for RGB star 42241. In
Table. \ref{table_compare} we provide
a comparison between $W_{\lambda}$s as measured from our observed
spectrum, and $W_{\lambda}$s as measured by us from the spectrum obtained by \citet{2008ApJ...688L..13K}.
We note that, for the Ca\,{\sc i}
lines, the spectrum from \citet{2008ApJ...688L..13K} has
deeper absorption. However, the overall absorption for the Fe\,{\sc i} and blended lines are in good
agreement, except for one weak Fe\,{\sc i} line at $\lambda=6200.31 \, \AA$ that is more prominent
in our observed spectrum. We note that the S/N at this line in the spectrum 
from \citet{2008ApJ...688L..13K} is low, making it difficult to distinguish such a
weak line in the spectrum.
There is a much
brighter star, SDSS J163056.63+124737.5, located only $\sim 12$ arcsec
from 42241. Thus, we investigate the possibility that the fibre
allocated for 42241 has collected a significant amount of flux from
SDSS J163056.63+124737.5. SDSS J163056.63+124737.5 is 6.7 magnitudes
brighter in the SDDS $r$-filter, which is centred on our wavelength
region of interest. The seeing for our observations was $\sim 1$
arcsec. Based on this, we constructed a model of two Gaussian flux
distributions with a full-width half-maximum of 1.5 arcsec and
magnitudes that represent the brightness of the stars. We found that
the amount of flux from SDSS J163056.63+124737.5 at the position of
the fibre is negligible. A similar investigation was carried out for the 
spectrum obtained by \citet{2008ApJ...688L..13K}, yielding the
same conclusion.
Thus, the origin of this discrepancy can not be due to a
contamination by light from this nearby star.
A more thorough investigation than this goes beyond the scope of this study.
However, one could speculate that there is an unresolved binary present that was
overlapped in one observation and out of phase in the other observation, or that it may be due to
some differences in the reduction procedure.
\\

\begin{figure}
\resizebox{\hsize}{!}{\includegraphics{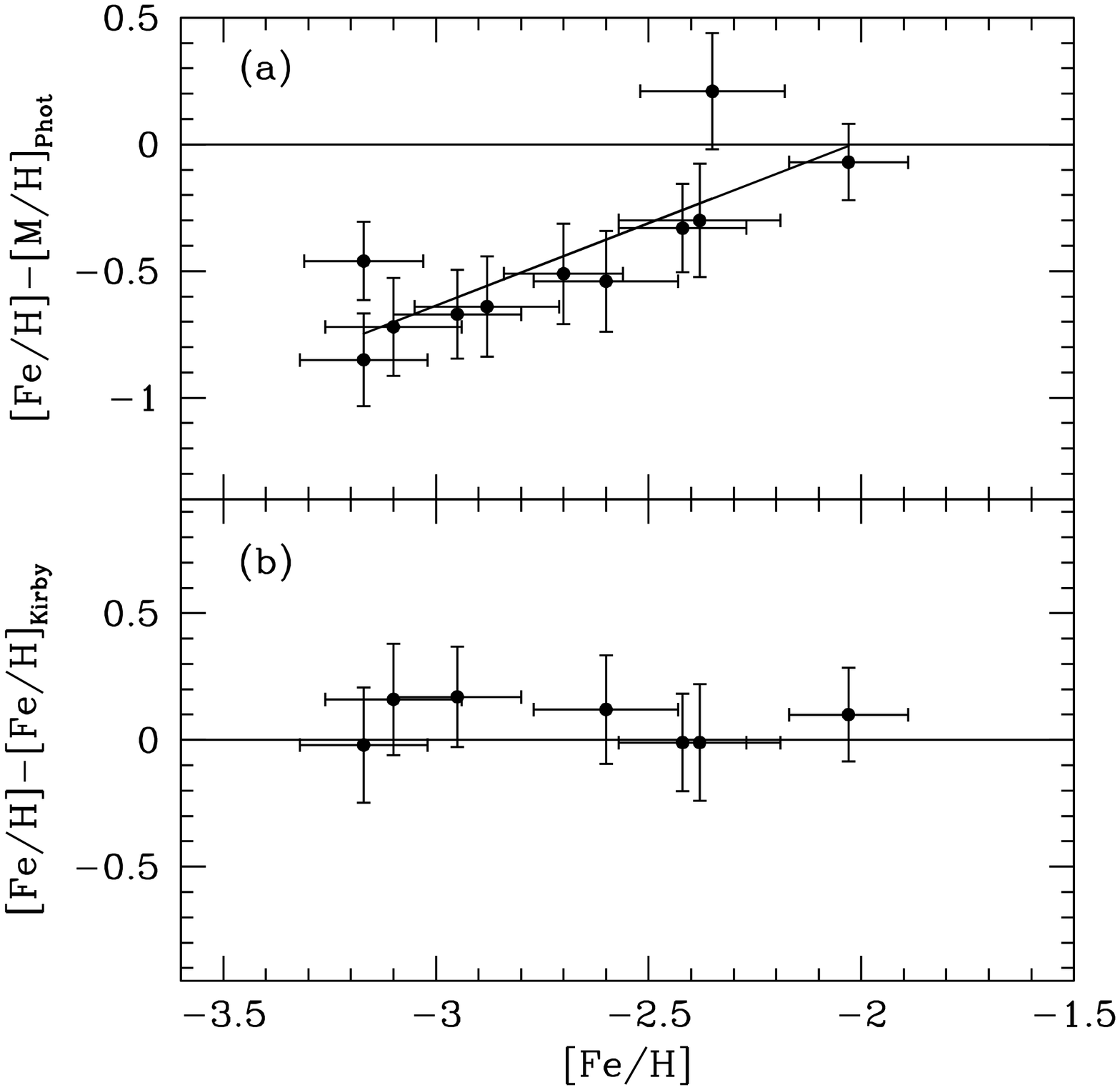}}
\caption{ {\bf a)} A comparison between our [Fe/H] and [M/H] from
  \citet{2009A&A...506.1147A} (${\rm [M/H]_{Phot}}$). The error-bars represent the error in ${\rm [Fe/H]-[M/H]_{Phot}}$
  and [Fe/H], respectively. {\bf b)} A
  comparison between our [Fe/H] and [Fe/H] from
  \citet{2008ApJ...685L..43K} (${\rm [Fe/H]_{Kirby}}$).
  The error-bars represent the error in ${\rm [Fe/H]-[Fe/H]_{Kirby}}$
  and [Fe/H], respectively.}
\label{fehcomp}
\end{figure}

\citet{2008ApJ...685L..43K} studied 20 stars in the direction of the
Hercules dSph galaxy. Their metallicities are based on a recently
developed automated spectrum synthesis method that takes the
information in the whole spectrum into account
\citep{2008ApJ...682.1217K}. The method was originally developed for
globular clusters in the Milky Way and was then applied to ultra-faint
dSph galaxies in \citet{2008ApJ...685L..43K}. We have 7 stars
in common between our samples.  Figure \ref{fehcomp}b shows the
difference between our respective determinations of [Fe/H].  We find that our
[Fe/H] is on average 0.07 dex more metal-rich, with a scatter of 0.09
dex.  In conclusion, the agreement between the [Fe/H] determinations is
very good.

\subsection{A comparison with earlier photometric results} \label{comparison_phot}

In a previous study of the Hercules dSph galaxy
\citep{2009A&A...506.1147A} we estimated [Fe/H] using the Str\"omgren
$m_1$ index using the calibration from \citet{2007ApJ...670..400C}. Figure \ref{fehcomp}a shows a comparison
between the photometric [Fe/H] as estimated in \citet{2009A&A...506.1147A}, ${\rm [M/H]_{phot}}$, and [Fe/H]
as determined from high-resolution spectroscopy in this study. We note that there is a strong trend such
that [M/H] appears to be over-estimated in
\citet{2009A&A...506.1147A} for metal-poor stars.

\subsection{A new metallicity calibration for Str\"omgren photometry for metal-poor red giant stars} \label{comparison_calib}

Given the excellent agreement between all three spectroscopic studies
it must be concluded that the metallicity calibration by \citet{2007ApJ...670..400C}
severely over-estimates the metallicity for very metal-poor stars. A check shows that also their updated
calibration \citep{2009ApJ...706.1277C} has the same problem.
This is an unfortunate situation since the photometry allows us in principle to determine
the metallicity of RGB stars with good accuracy also for the fainter
stars \citep[compare errors in][]{2009A&A...506.1147A} and thus allowing
the study of much more complete stellar samples in the ultra-faint dSph galaxies.

Here we present an attempt to deal with the situation. So far this is a very simplistic relation and
{\em only} formally valid for stars with $0.02<m_{1,0}<0.40$, $-3.29<{\rm [Fe/H]}<1.58$ and 
$1.15<(v-y)_0<2.18$.

\begin{figure}
\resizebox{\hsize}{!}{\includegraphics{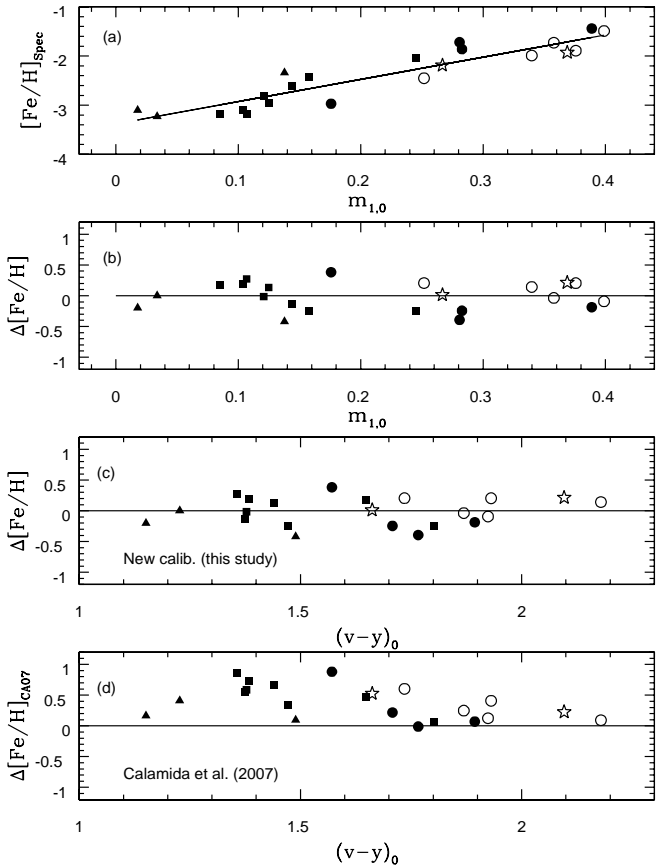}}
\caption{{\bf a)} Spectroscopic [Fe/H] vs. $m_{1,0}$ for Draco ($\bullet$ and $\circ$), Sextans (open stars),
UMaII (filled triangles) and Hercules (filled squares). {\bf b)}, ${\rm [M/H]_{phot, new} - [Fe/H]}$ vs. $m_{1,0}$ using Eq. (\ref{eq_phot_new}) to derive [M/H]. {\bf c)}, ${\rm [M/H]_{phot, new} - [Fe/H]}$ vs. $(v-y)_0$ using Eq. (\ref{eq_phot_new}) to derive [M/H]. {\bf d)} ${\rm [M/H]_{CA07} - [Fe/H]}$ vs. $(v-y)_0$ using the calibration by \citet{2007ApJ...670..400C} to derive [M/H].}
\label{calib2}
\end{figure}

We have collected all spectroscopic [Fe/H] derived from high-resolution spectra
available for stars in Draco \citep{2009ApJ...701.1053C,2001ApJ...548..592S}, Sextans \citep{2001ApJ...548..592S}, UMaII \citep{2010ApJ...708..560F} and Hercules (this study), and combined these data with our own
Str\"omgren photometry where available. Fig. \ref{calib2}a shows the spectroscopic [Fe/H] as a function of $m_{1,0}$ for the stars.
A least-squares fit yields

\begin{equation} \label{eq_phot_new}
{\rm [M/H]_{phot, new}}=4.51(\pm 0.41) \cdot m_{1,0} - 3.38(\pm 0.10)
\end{equation}

Fig. \ref{calib2}b and \ref{calib2}c show ${\rm [M/H]_{phot, new} - [Fe/H]}$ as a function of $m_{1,0}$ and $(v-y)_0$, respectively.
No strong trends are seen. For comparison Fig. \ref{calib2}d shows ${\rm [M/H]_{CA07} - [Fe/H]}$ vs. $(v-y)_0$, where
${\rm [M/H]_{CA07}}$ is [Fe/H] as determined using the calibration in
\citet{2007ApJ...670..400C}. Here we can note a significant difference of both metallicity scales
such that ${\rm [M/H]_{CA07}}>{\rm [Fe/H]_{spec}}$.
Taking into account the uncertainties of the least-squares fit and the correlation between the fitting parameters (Eq. \ref{eq_phot_new}), and the
 error in $m_{1,0}$ from \citet{2009A&A...506.1147A} we find a typical error in ${\rm [M/H]_{phot, new}}$
 of 0.17 dex.

\section{Results and discussion} \label{discu}

\subsection{Iron abundance and ages for the RGB stars in Hercules}

The RGB stars analysed in this paper span a large range of iron 
abundances, from about --3.2\,dex to --2\,dex, indicating an extended
period of chemical enrichment. It is somewhat fortuitous that the
two brightest stars in our sample, RGB stars 12175 and 42241, bracket
the full range of metallicities. Thus there is no doubt
that the range of metallicities derived from high-resolution
spectroscopy is real.

In Sect. \ref{comparison_calib} we provide a new Str\"omgren metallicity calibration.
This calibration is
valid for stars with $0.02<m_{1,0}<0.40$ and $-3.29<{\rm [Fe/H]}<1.58$.
Two of the 28 RGB stars from \citet{2009A&A...506.1147A} have an $m_{1,0}$ less than the range for which
the new metallicity calibration is valid.
However, with an $m_{1,0}$ of 0.01, these stars are included in the sample as a slight extrapolation.
In Fig. \ref{new_calib2}a we show the
resulting histogram of ${\rm [M/H]_{phot, new}}$ for all the 28 RGB stars identified in \citep{2009A&A...506.1147A}.
The bin size of 0.2 dex represents the typical error in ${\rm [M/H]_{phot, new}}$ (see Sect. \ref{comparison_calib}). 
{Figure  \ref{new_calib2}b shows the corresponding error-weighted metallicity distribution. For this plot, each stars was assigned a Gaussian distribution with a mean of ${\rm [M/H]_{phot, new}}$ and a dispersion equal to the typical error in ${\rm [M/H]_{phot, new}}$ (0.17 dex). The Gaussians, one for each star, were then added to create the metallicity distribution function.
We note that the distribution of ${\rm [M/H]_{phot, new}}$ is shifted towards lower metallicities
when the new calibration is applied, and that there is an abundance spread in the metallicity distribution
for the RGB stars of at least 1.0 dex. Additionally, we note a more concentrated distribution.

\begin{figure}
\resizebox{\hsize}{!}{\includegraphics{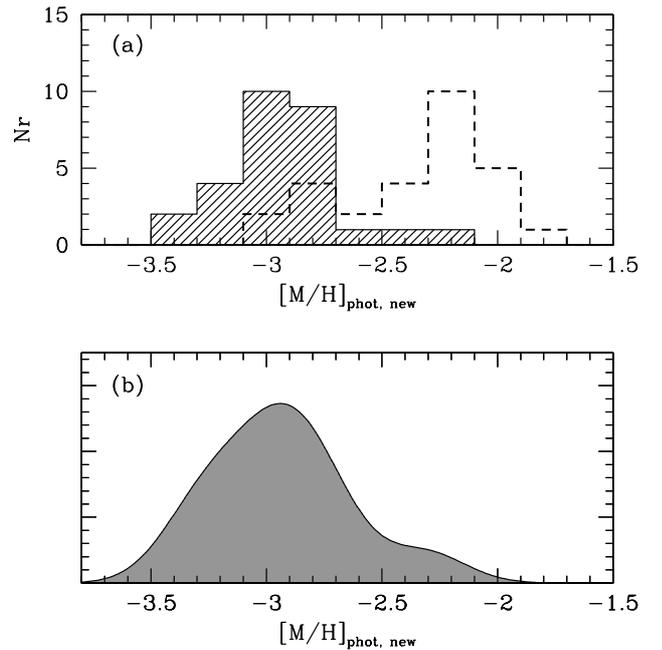}} 
\caption{{\bf a)} Metallicity histogram for RGB stars in the Hercules dSph galaxy. The
shaded histogram shows the distribution of ${\rm [M/H]_{phot, new}}$. For comparison, the dashed
histogram shows the distribution of ${\rm [M/H]_{phot}}$ \citep{2009A&A...506.1147A}.
 {\bf b)} Corresponding error-weighted metallicity distribution.}
\label{new_calib2}
\end{figure}

Figure\,\ref{iso2}a and \ref{iso2}b show $V_0$ vs. $(v-y)_0$ for the 
stars with [Fe/H] derived from high resolution spectroscopy.
Additionally, in these plots, we show two isochrones with 
[Fe/H]$=-2.31$ (most metal-poor isochrone available) and $-2.14$ dex.
As can be seen, the isochrones of a given metallicity become
redder with increasing age. Since the isochrone with [Fe/H]$=-2.31$ is too
metal-rich, compared to [Fe/H] as derived from the spectroscopy, an even
more metal-poor isochrone at the age of 8 Gyr would be even bluer, excluding an age
of about 10 Gyr or younger. At an age of 14 Gyr, the isochrone with [Fe/H]$=-2.31$ is slightly
redder than most of the stars more metal-poor than [Fe/H]$=-2.7$. Hence
a more metal-poor isochrone would presumably represent the locus of these
stars very well, arguing for an age older than about 10 Gyr for the Hercules dSph galaxy.

\begin{figure}
\resizebox{\hsize}{!}{\includegraphics{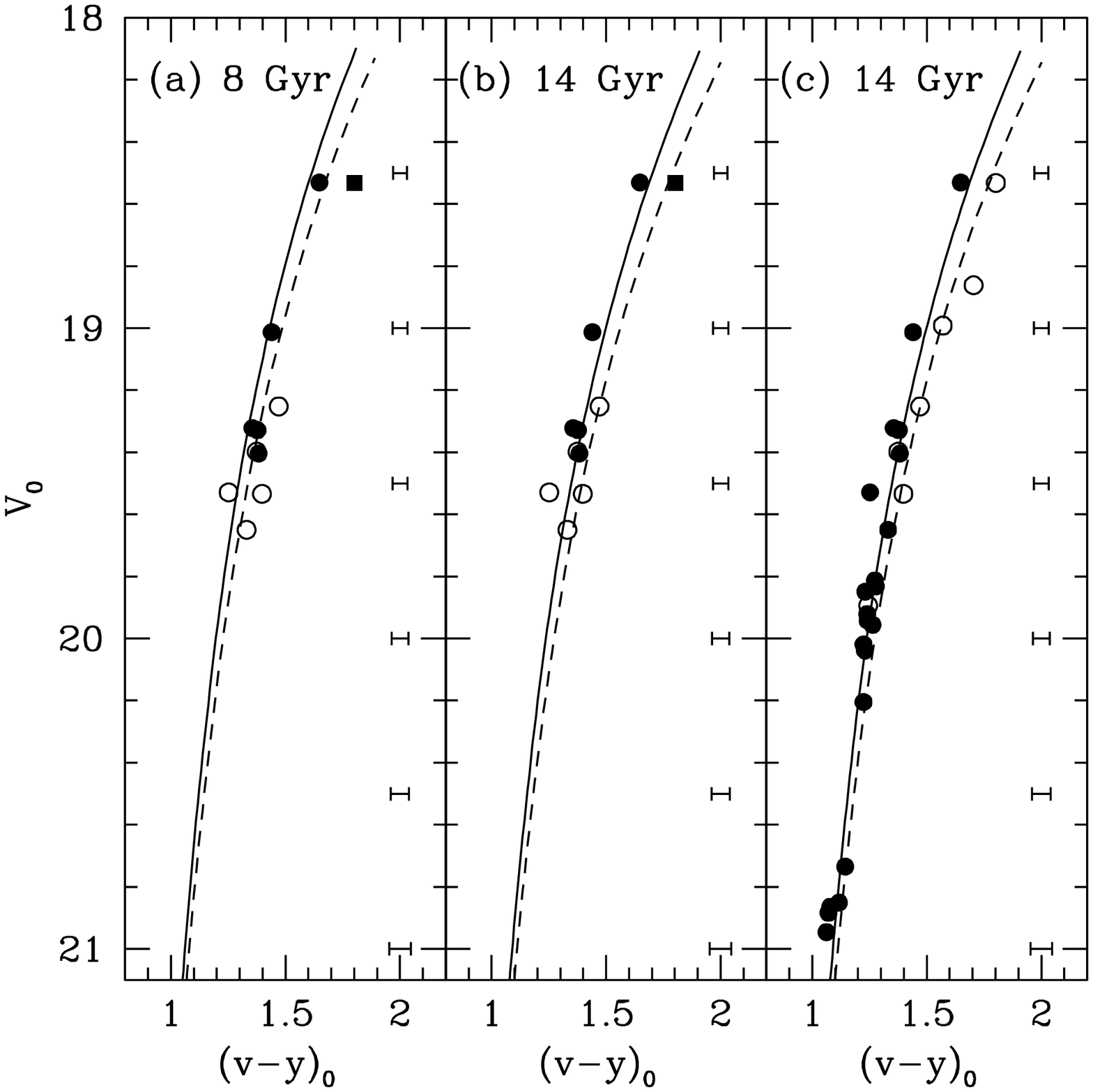}}
\caption{ {\bf a)} and {\bf b)} Colour-magnitude diagrams for RGB stars with
high-resolution spectroscopy of the Hercules dSph galaxy.
  $\bullet$ indicate RGB stars more metal-poor than [Fe/H]$=-2.7$.
  $\circ$ represents RGB stars more metal-rich than, or equal to, [Fe/H]$=-2.7$. The filled
  square indicates the most metal-rich RGB star at [Fe/H]$=-2.0$.
  The solid and dashed lines represent isochrones with [Fe/H]$=-2.31$ and $-2.14$ dex, respectively,  by
  \citet{2006ApJS..162..375V} with colour transformations by
  \citet{2004AJ....127.1227C}. {\bf c)} Colour-magnitude diagram for the RGB stars 
 in \citet{2009A&A...506.1147A} with ${\rm [M/H]_{phot, new}}$ as determined in Sect. \ref{comparison_calib}.
  $\bullet$ indicate RGB stars more metal-poor than [M/H]$=-2.8$.
  $\circ$ indicate RGB stars more metal-rich than [M/H]$=-2.8$.
  Isochrones as in ({\bf b}). The error bars on the right hand side in each figure 
  represent the typical error in $(v-y)_0$.}
\label{iso2}
\end{figure}

Figure \ref{iso2}c shows $V_0$ vs. $(v-y)_0$ for all the 28 RGB stars identified in \citep{2009A&A...506.1147A}.

\subsection{[Ca/Fe]}

Figure \ref{cafe_feh} shows [Ca/Fe] as a function of [Fe/H].  We find
a trend such that [Ca/Fe] is higher for more metal-poor stars, and
lower for more metal-rich stars. Fortuitously, the most metal-rich and
the most metal-poor star in the sample are both bright and have
spectra with high S/N (see discussion in Sect.\,\ref{abund} and also
Table\,\ref{table_obs}). Thus we can be certain that the trend
actually has this shape and we are not misinterpreting spectra of
lower quality.

\begin{figure}
\resizebox{\hsize}{!}{\includegraphics{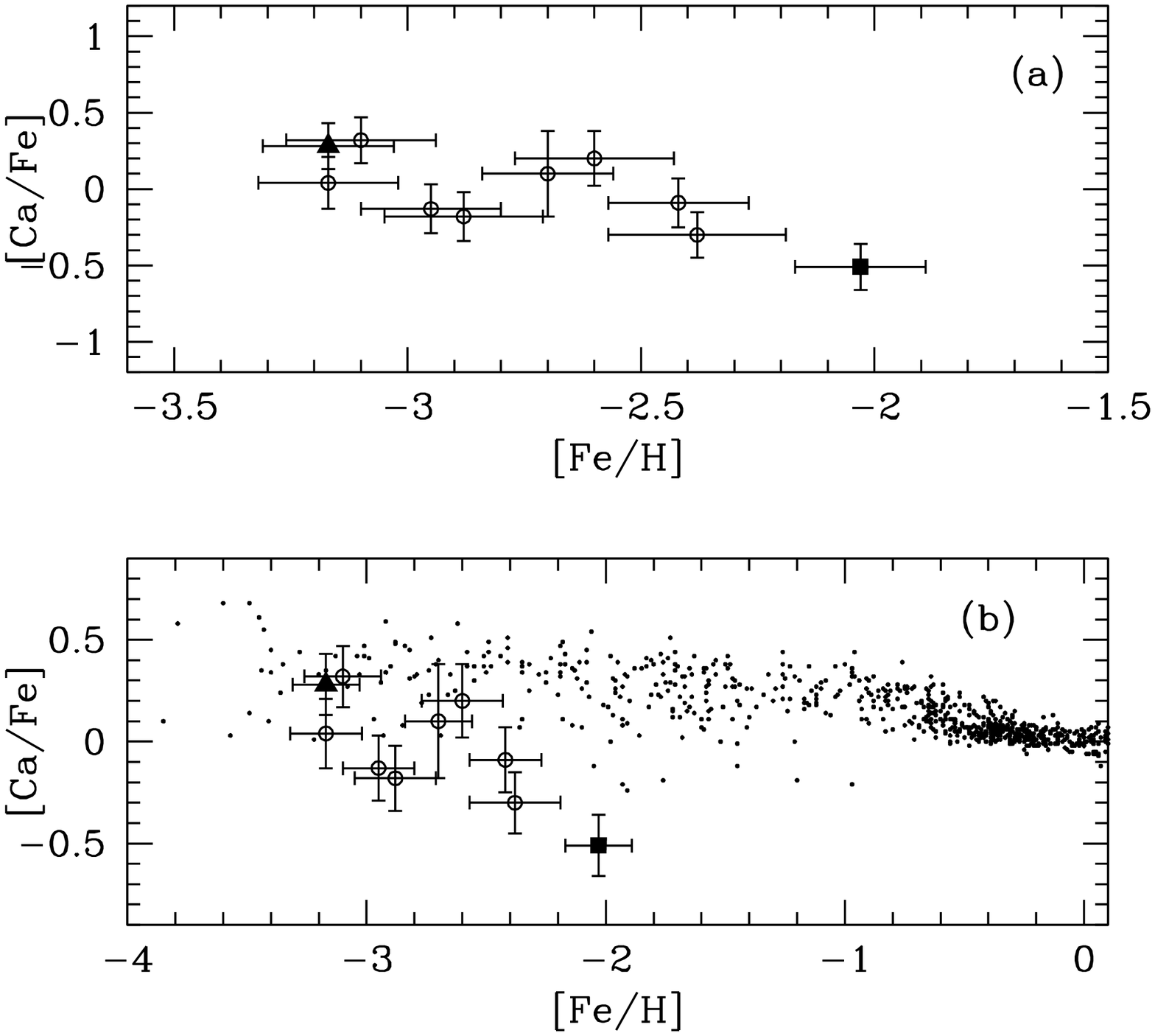}} 
\caption{{\bf a)} [Ca/Fe] as a function of [Fe/H]. Filled triangle and filled
  square indicate our two brightest RGB stars, RGB stars 12175 and
  42241, respectively.  The error-bars represent the error in [Ca/Fe]
  and [Fe/H], respectively. {\bf b)} Same as {\bf a} but with a compilation of
  the Milky Way disk and halo stars abundances from \citet{2004AJ....128.1177V}, as indicated by small dots.}
\label{cafe_feh}
\end{figure}

The production of alpha ($\alpha$)-elements, such as Ca, Si, Ti, Mg,
and O, is correlated with the end stage of massive stars. Mg and O are
created during the hydrostatic He burning in massive stars, and Si, Ca,
and Ti are primarily produced during core-collapse supernovae \citep{1995ApJS..101..181W}.  
On the other hand, less
massive stars are able to produce significant amounts of Fe in SNe\,Ia.
Thus, the ratio of $\alpha$-elements to iron is used to trace the time
scale of the star formation in a stellar system. If the star formation rate is high,
then the gas will be able to reach a higher [Fe/H] before the first SNe\,Ia occur.
This can be observed in a plot of [Ca/Fe] vs. [Fe/H] as a "knee",
where [Ca/Fe] decrease as [Fe/H] increase \citep{1997ARA&A..35..503M}.
The fraction of stars at [Fe/H] less than the "knee" gives information on the star formation
timescale.

The observed continuous downward trend, without a "knee", for [Ca/Fe] vs. [Fe/H] in Hercules can
thus be interpreted as a brief initial burst of short-lived SNe\,II that enhanced
the production of $\alpha$-elements. Since there are no stars at [Fe/H] less than the "knee",
the star formation rate was very low. The subsequent continuous decline
would be expected if contributions from long-lived SNe\,Ia were the dominant factor,
decreasing [$\alpha$/Fe] while increasing [Fe/H]. This means that essentially
no massive stars formed after the initial burst. Additionally, we interpret
the relatively short range in [Fe/H] (no stars with [Fe/H]$>-2$ are seen in our sample) as a tentative evidence for a short
duration of this low-efficiency star formation \citep[for a discussion of
continuous and bursty star formation histories and the role of SNe\,Ia
see, e.g.,][]{1991ApJ...367L..55G,2009MmSAI..80...83M}. The classical
dSph galaxies, such as Carina, Sculptor and Fornax, also show these types of trends for the
$\alpha$-elements \citep[e.g.,][]{2004AJ....128.1177V,2008AJ....135.1580K,2009ARA&A..47..371T,
2009ApJ...705..328K}. However,
these dSph galaxies are more metal-rich and more massive than, e.g.,
Hercules. Only a few other ultra-faint dSphs have chemical element abundances
published for only a handful of stars each.

\citet{2010ApJ...708..560F}, \citet{2009A&A...508L...1F}, \citet{2010ApJ...711..350N} and \citet{2010arXiv1001.3137S} studied Coma
Berenices, Ursa Major\,II, Bo\"otes\,I, and Leo IV, all recently discovered
ultra-faint and metal-poor systems. Figure \ref{cafe_feh_other} summarises our data
and their data.
Additionally, recent studies have analysed very metal-poor stars in the
classical systems Draco, Sextans and Sculptor
\citep{2009ApJ...701.1053C,2009A&A...502..569A,2010Natur.464...72F}. We add these new data
to the plot in addition to the abundances from other studies
of Fornax, Carina, Sculptor, Sextans, Ursa Minor and Draco
\citep{2001ApJ...548..592S,2003AJ....125..684S,2004PASJ...56.1041S,2007EAS....24...33L,2008AJ....135.1580K}.

\begin{figure*}
\resizebox{\hsize}{!}{\includegraphics{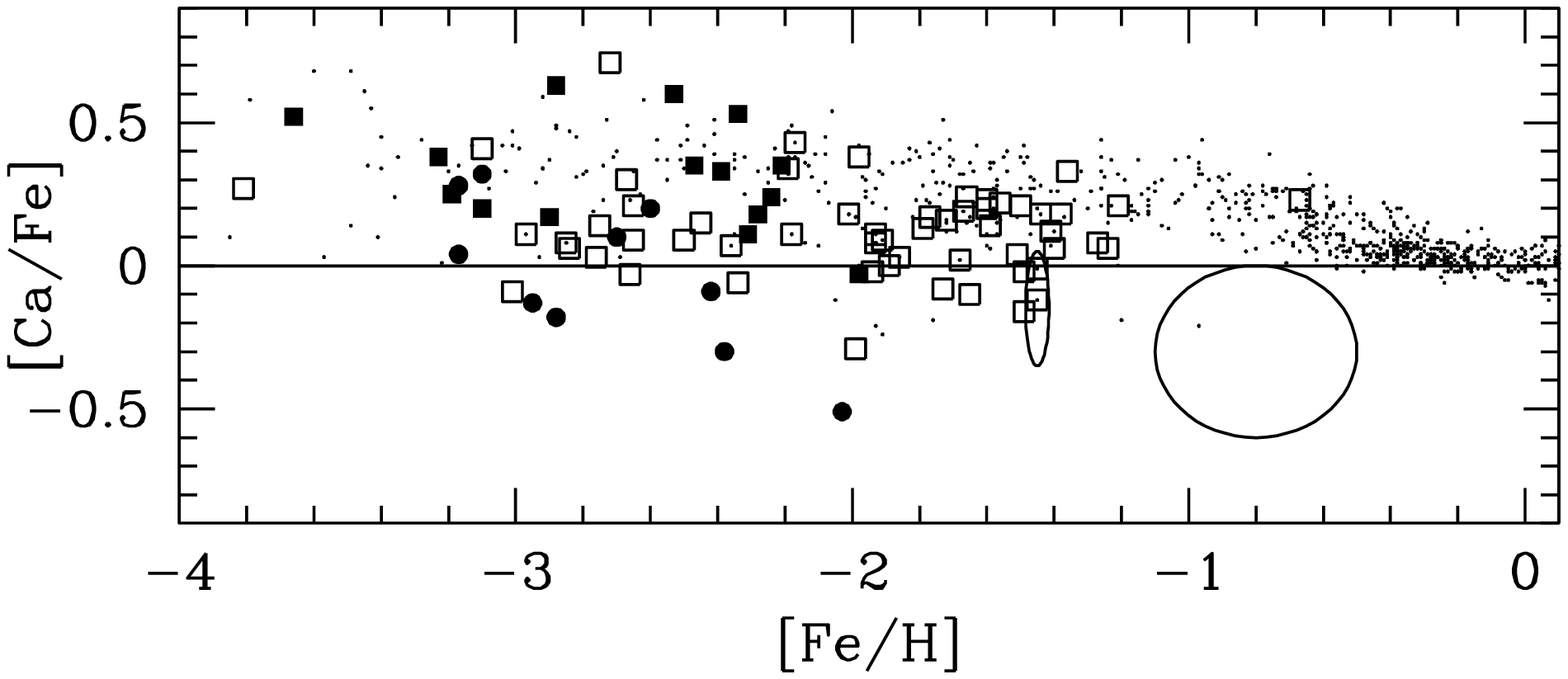}} 
\caption{A comparison of [Ca/Fe] as a function of [Fe/H] for stars in
  several dSph galaxies. $\bullet$ indicate Hercules (this study). Filled squares represent
  the ultra-faint dSph galaxies Ursa Major II, Coma Berenices, Bo\"otes I and Leo IV
  \citep{2009A&A...508L...1F,2010ApJ...708..560F,2010ApJ...711..350N,2010arXiv1001.3137S}.
  Open squares represent the classical dSph galaxies Draco, Sextans, Ursa Minor, Fornax,
  Carina and Sculptor \citep{2001ApJ...548..592S,2003AJ....125..684S,2004PASJ...56.1041S,2008AJ....135.1580K,2009ApJ...701.1053C,2009A&A...502..569A,2010Natur.464...72F}.
  The solid ellipses outline RGB stars in the classical dSph galaxy Fornax from \citet{2007EAS....24...33L}.
  }
\label{cafe_feh_other}
\end{figure*}

Overall, there is a faster declining [Ca/Fe] with [Fe/H]
for the dSph galaxies as compared with the halo stars in the solar neighbourhood 
\citep[from the compilation by][including data from \citet{2002AJ....123..404F,2000AJ....120.1841F,2002AJ....123.1647S,2003A&A...410..527B,1997A&A...326..751N,1998AJ....116.1286H,2000AJ....120.2513P,2003MNRAS.340..304R,1993A&A...275..101E,1998AJ....115.1640M,1995AJ....109.2736M,2002ApJS..139..219J,2000ApJ...544..302B,2003ApJ...592..906I,1996ApJ...471..254R,1991A&A...241..501G,1994A&A...287..927G,1988A&A...204..193G}]{2004AJ....128.1177V}.
Thus, for example, the trend
seen from our data in Hercules is the same as the overall trend seen
for Draco. This is interesting and could be interpreted as that the Fe contribution from 
SNe\,Ia were the dominant factor for both Hercules and Draco. However, since Draco
has many more stars with [Fe/H]$> -2$ it must have had a more integrated star formation 
than the Hercules dSph galaxy, as may be expected given its higher baryon content.

\section{Conclusions} \label{conclusion}

We have studied confirmed RGB stars in the ultra-faint Hercules dSph galaxy with FLAMES
high-resolution spectroscopy. Abundances were determined by solving the radiative transfer
calculations using the codes Eqwi/Bsyn in {\sc marcs} model atmospheres.

We find that the RGB stars of the Hercules dSph galaxy included in this study are more metal-poor than
estimated in \citet{2009A&A...506.1147A}, however in good agreement with \citet{2008ApJ...685L..43K}, with a metallicity spread of at least 1 dex.
Based on the position of the RGB stars in colour-magnitude diagrams, in comparison with isochrones, we conclude that there is no clear indication of a population younger than about 10 Gyr.

Additionally, we provide a first attempt at a new metallicity calibration for Str\"omgren photometry
based on high-resolution spectroscopy for several dSph galaxies. With this new calibration, we find
several RGB stars in the Hercules dSph galaxy that are more metal-poor than [Fe/H]=--3.0 dex.

Finally, we have determined the [Ca/Fe] for the RGB stars in this study. We found a trend such that
[Ca/Fe] is higher for more metal-poor stars, and lower for more metal-rich stars. This trend
is supported by our two brightest stars in the sample and is interpreted as a brief initial burst
of SNe\,II during a very low star formation rate, followed by the enrichment of [Fe/H] by SNe\,Ia.

\begin{acknowledgements}
We acknowledge Karin Lind for providing us with a spectrum of one of their RGB stars.
S.F. is a Royal Swedish Academy of Sciences Research Fellow supported by a grant from the 
Knut and Alice Wallenberg Foundation.
K.E is gratefully acknowledging support from the Swedish
research council.
M.I.W. is supported by a Royal Society University Research Fellowship. 
AK acknowledges support by an STFC postdoctoral fellowship.

\end{acknowledgements}

\bibliographystyle{aa} 
\bibliography{herc_hr}

\end{document}